\documentclass[sn-mathphys,iicol]{sn-jnl}% Math and Physical Sciences Reference Style
\usepackage{siunitx}
\jyear{2022}%

\begin{document}

\title[ ]{\vspace{-1.5cm}Muonic atom spectroscopy with microgram target material}

\author[1]{\fnm{A.} \sur{Adamczak}} % confirmed
\author[2,3]{\fnm{A.} \sur{Antognini}} % confirmed
\author[4,5]{\fnm{N.} \sur{Berger}} % confirmed
\author[6]{\fnm{T. E.} \sur{Cocolios}} % confirmed
\author[4,5]{\fnm{N.} \sur{Deokar}} % confirmed
\author[5,7,8,9]{\fnm{Ch. E.} \sur{D\"ullmann}} % confirmed
\author[3]{\fnm{A.} \sur{Eggenberger}} % confirmed
\author[2]{\fnm{R.} \sur{Eichler}} % confirmed
\author[6]{\fnm{M.} \sur{Heines}} % confirmed
\author[10]{\fnm{H.} \sur{Hess}} % confirmed
\author[11]{\fnm{P.} \sur{Indelicato$^{\textrm{11}}$}} % confirmed
\author[2,3]{\fnm{K.} \sur{Kirch}} % confirmed
\author*[2]{\fnm{A.} \sur{Knecht}}\email{a.knecht@psi.ch} % confirmed
\author[5,12]{\fnm{J.J.} \sur{Krauth}} % confirmed
\author[2,3]{\fnm{J.} \sur{Nuber}} % confirmed
\author[5,12]{\fnm{A.} \sur{Ouf}} % confirmed
\author[2,13]{\fnm{A.} \sur{Papa}} % confirmed
\author[5,12]{\fnm{R.} \sur{Pohl}} % confirmed
\author[2]{\fnm{E.} \sur{Rapisarda}} % confirmed
\author[10]{\fnm{P.} \sur{Reiter}} % confirmed
\author[2,3]{\fnm{N.} \sur{Ritjoho}} % confirmed
\author[14]{\fnm{S.} \sur{Roccia}} % confirmed
\author[10]{\fnm{M.} \sur{Seidlitz}} % confirmed
\author[6]{\fnm{N.} \sur{Severijns}} % confirmed
\author[3]{\fnm{K.} \sur{von Schoeler}} % confirmed
\author[2,3]{\fnm{A.} \sur{Skawran}} % confirmed
\author[2,3]{\fnm{S. M.} \sur{Vogiatzi}} % confirmed
\author[10]{\fnm{N.} \sur{Warr}} % confirmed
\author[4,5]{\fnm{F.} \sur{Wauters}} % confirmed

\affil[1]{\orgname{Institute of Nuclear Physics, Polish Academy of Sciences}, \city{Krakow}, \country{Poland}}
\affil[2]{\orgname{Paul Scherrer Institut}, \city{Villigen}, \country{Switzerland}}
\affil[3]{\orgdiv{Institut f\"ur Teilchen- und Astrophysik}, \orgname{ETH Z\"urich}, \city{Z\"urich}, \country{Switzerland}}
\affil[4]{\orgdiv{Institute of Nuclear Physics}, \orgname{Johannes Gutenberg University Mainz}, \city{Mainz}, \country{Germany}}
\affil[5]{\orgdiv{PRISMA+ Cluster of Excellence}, \orgname{Johannes Gutenberg University Mainz},  \city{Mainz} \country{Germany}}
\affil[6]{\orgname{KU Leuven}, \orgdiv{Instituut voor Kern- en Stralingfysica}, \city{Leuven}, \country{Belgium}}
\affil[7]{\orgdiv{Department of Chemistry - TRIGA Site}, \orgname{Johannes Gutenberg University Mainz}, \city{Mainz}, \country{Germany}}
\affil[8]{\orgname{GSI Helmholtzzentrum f\"ur Schwerionenforschung}, \city{Darmstadt}, \country{Germany}}
\affil[9]{\orgname{Helmholtz Institute Mainz}, \city{Mainz}, \country{Germany}}
\affil[10]{\orgdiv{Institut f\"ur Kernphysik}, \orgname{Universit\"at zu K\"oln}, \city{K\"oln}, \country{Germany}}
\affil[11]{\orgdiv{Laboratoire Kastler Brossel}, \orgname{Sorbonne Universit\'e, CNRS, ENS-PSL Research University, Coll\`ege de France, Case\ 74;\ 4, place Jussieu}, \city{F-75005 Paris}, \country{France}}
\affil[12]{\orgdiv{Institute of Physics}, \orgname{Johannes Gutenberg Universit\"at Mainz},  \city{Mainz}, \country{Germany}}
\affil[13]{\orgdiv{Department of Physics}, \orgname{Universit\'a di Pisa},  \city{Pisa}, \country{Italy}}
\affil[14]{\orgdiv{Universit\'e Grenoble Alpes}, \orgname{CNRS, Grenoble INP, LPSC-IN2P3},  \city{38026 Grenoble}, \country{France}}


\abstract{Muonic atom spectroscopy -- the measurement of the x rays emitted during the formation process of a muonic atom -- has a long standing history in probing the shape and size of nuclei. In fact, almost all stable elements have been subject to muonic atom spectroscopy measurements and the absolute charge radii extracted from these measurements typically offer the highest accuracy available. However, so far only targets of at least a few hundred milligram could be used as it required to stop a muon beam directly in the target to form the muonic atom. We have developed a new method relying on repeated transfer reactions taking place inside a 100~bar hydrogen gas cell with an admixture of 0.25\% deuterium that allows us to drastically reduce the amount of target material needed while still offering an adequate efficiency. Detailed simulations of the transfer reactions match the measured data, suggesting good understanding of the processes taking place inside the gas mixture. As a proof of principle we demonstrate the method with a measurement of the 2\textit{p}-1\textit{s} muonic x rays from a \SI{5}{\micro\gram} gold target.}




\maketitle

\section{Introduction}

Muonic atoms are a type of exotic atoms where a negative muon ($\mu^-$) is captured by an atomic nucleus and forms a hydrogen-like system. For this capture process to take place, muons need to be decelerated to a few eV, where they are then captured in a high Rydberg state (main quantum number $n\geq$ 14). From this state they subsequently cascade down to the $1s$ state via collisions and Auger and radiative transitions, during the latter step emitting so-called muonic x rays. The details of this capture and cascade depend on the target material and density~\cite{STANISLAUS1987642,Markushin:1998ve}, the entire process transpires however within at most a few nanoseconds for low-$Z$ materials\footnote{Except for transitions passing through the $2s$ state, which can be metastable.} (much less for higher-$Z$ elements), which is less than the time resolution of a typical x-ray detector. Once a muonic atom is formed, the muon is unlikely to transfer to another atom, with the exception of muonic hydrogen which is a neutral system and thus can easily penetrate the electron cloud of nearby atoms. In the ground state, the muon will either decay in orbit, or be captured by the atomic nucleus~\cite{Suzuki:1987jf}. This process becomes dominant for $Z$$\geq$14 due to a strong $Z^4$ dependence, thereby considerably reducing the lifetime of the exotic atom.

Due to the relatively large muon mass of 105.7~MeV/c$^2$, the muonic Bohr radius is $\sim$207 times smaller compared to regular atoms leading to a large overlap between the muonic and nuclear wave function at lower $n$ orbits. This makes muonic atoms a very sensitive system to probe the nuclear charge distribution, first and foremost the nuclear charge radius $\sqrt{<r^2>}$ which is typically derived from the $2p-1s$ transition energy~\cite{Fri95,Knecht:2020npz}. In this way muonic atoms provide complementary information to electron scattering and laser spectroscopy data, respectively~\cite{ANGELI201369,agmb2009,Frois:1987hk,Cheal12,Neugart:2017oqg}.

To measure the characteristic muonic x rays from a particular isotope, one needs to stop a secondary muon beam available at a proton accelerator facility such as HIPA at the Paul Scherrer Institute (PSI), MUSE at J-PARC, the muon sources at TRIUMF and RAL, and MuSIC at Osaka. These muons have a typical momentum of 10-100 MeV/c. To efficiently stop such muons, one usually needs a solid target thickness of $\mathcal{O}$(1~mm)~\cite{GROOM2001183}, which limits the available muonic x-ray data to stable isotopes from which one can produce such a macroscopic target. However, the charge radii of several long lived radioactive isotopes are of interest to benchmark isotope shift measurements from laser spectroscopy~\cite{Cocolios:2717871}, or to serve as input for atomic parity violation experiments with radium~\cite{PhysRevA.78.050501}, which was the motivation to develop the method presented in this paper. In order to perform laser spectroscopy on the lightest muonic atoms, the Lamb Shift experiment at PSI has developed a dedicated low-energy muon beamline to slow down muons to a few keV in kinetic energy ~\cite{Antognini:2021icf}. In this paper we present an alternative approach, where a standard negative muon beam is stopped in a high-pressure gaseous hydrogen target, and the negative muon is then transported to a few \si{\micro\gram}/cm$^2$ target by means of diffusion and a series of transfer reactions. A total efficiency per incoming muon of order one percent is achieved in this way.

The main focus of the paper is the detailed description of the developed method (Section~\ref{sec_method}) and apparatus (Section~\ref{sec_apparatus}) allowing us to perform muonic atom spectroscopy with such microgram targets. Additionally, we have developed a Monte Carlo program to simulate all relevant processes taking place in the high-pressure gas cell. (Section~\ref{sec_simulation}). The paper concludes in Section~\ref{sec:results} with a description of the first results that show the performance of the apparatus in comparison to simulation and a first proof-of-principle measurement with a 5~\si{\micro\gram} gold target.

\begin{figure}%
\centering
\includegraphics[width=0.5\textwidth]{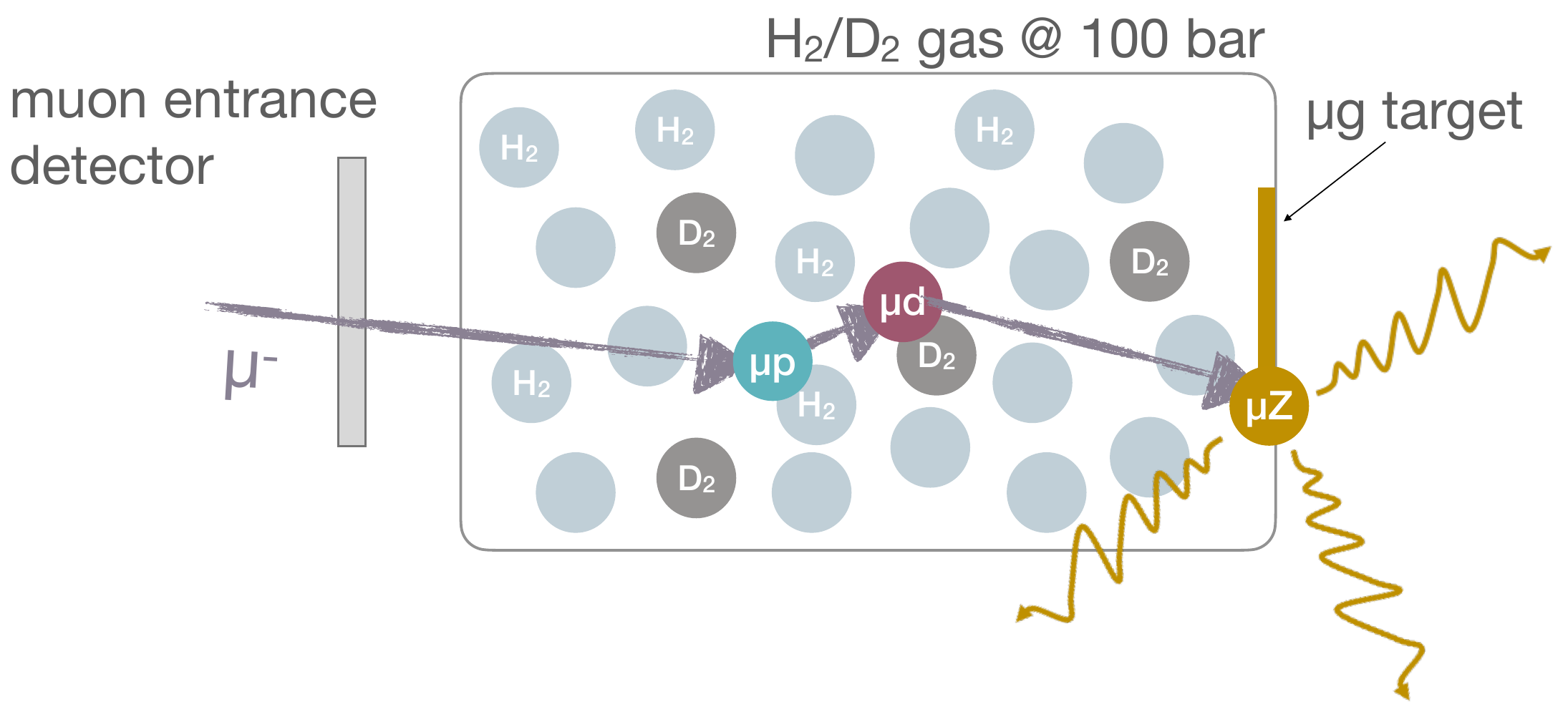}
\caption{Sketch of the method employed to measure muonic x rays emitted from a target with only microgram mass. The negative muon enters a 100~bar hydrogen gas cell with a small admixture of deuterium. Upon stopping, the muon forms muonic hydrogen $\mu p$ and transfers to a deuteron upon collision with a deuterium molecule forming muonic deuterium $\mu d$. Due to its low scattering cross section, $\mu d$ can travel over a large distance potentially reaching the microgram target at the back of the cell. Here, the muon transfers again to the target nucleus thereby emitting the characteristic muonic x rays.} \label{fig_sketch_gasCell}
\end{figure}

\section{Method}\label{sec_method}
The method presented in this paper for performing muonic atom spectroscopy with microgram target material relies on transfer reactions taking place inside a high-pressure hydrogen gas cell operating at 100~bar, corresponding to about 10\% of liquid hydrogen density, with a small admixture of deuterium. The method is inspired by the work described in Refs. \cite{kra89, str05, Str09} and the wealth of knowledge gained in the pursuit of muon catalyzed fusion (see, e.g., Refs. \cite{Pon90, pet01}) on the behavior and interaction of the muonic hydrogen isotopes inside gas cells. A sketch of the method can be found in Fig.~\ref{fig_sketch_gasCell} with the details given in the following paragraphs.

\begin{figure}%
\centering
\includegraphics[width=0.5\textwidth]{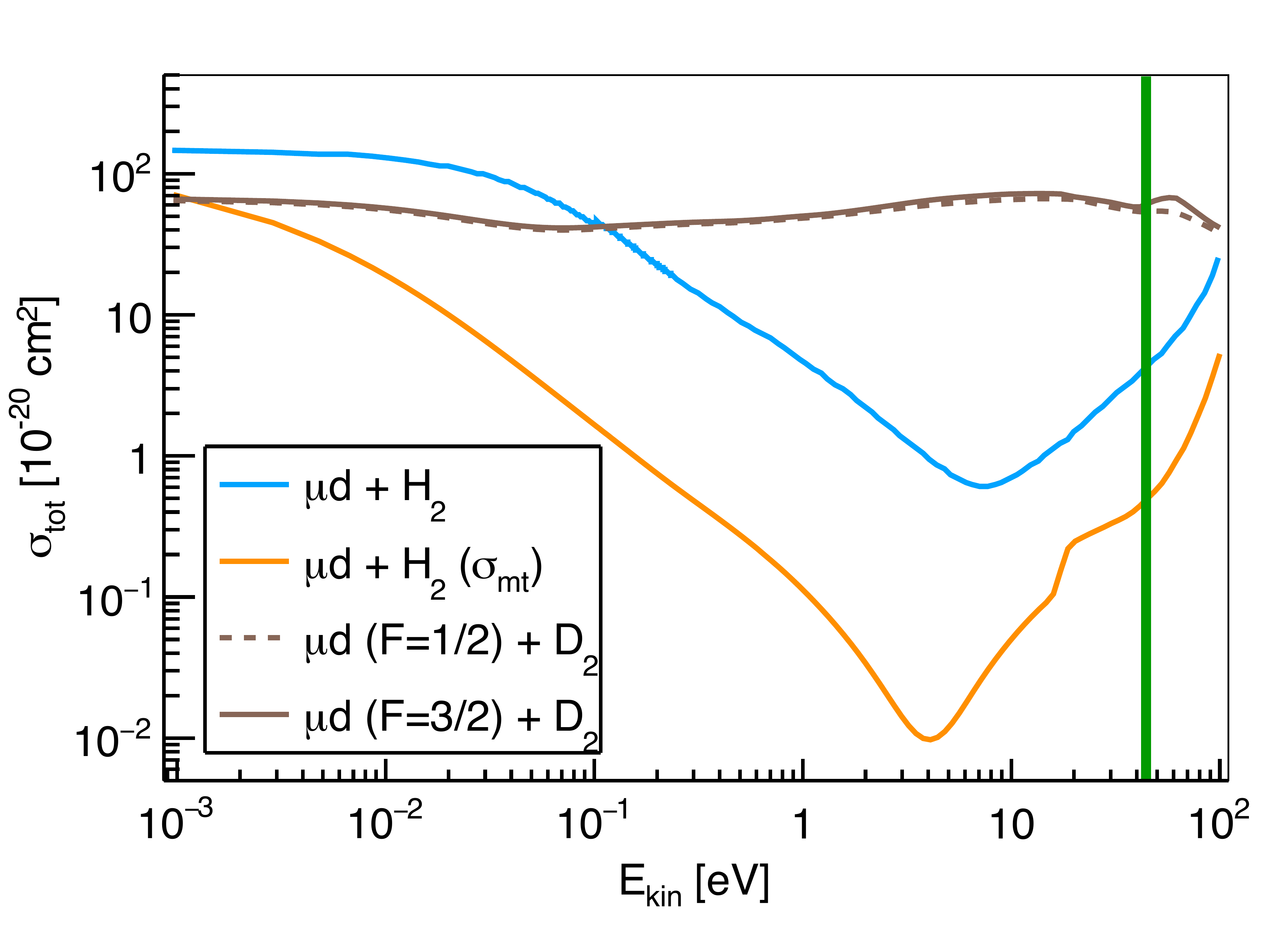}
\caption{Calculated total cross sections for the elastic scattering of $\mu d$ on H$_2$ and D$_2$ and transport cross sections (indicated with $\sigma_\text{mt}$) for the elastic scattering of $\mu d$ on H$_2$ \cite{Ada96}. The gain in kinetic energy by the transfer from $\mu p$ to $\mu d$ is indicated by the vertical green line. The cross sections for the scattering of $\mu d$ on D$_2$ depend on the hyperfine state $F$ of the $\mu d$ atom and are included in brown.} \label{fig_scattering-XS}
\end{figure}

Once the negative muon comes to rest inside the hydrogen gas it is quickly captured by a hydrogen molecule forming the excited $\mu$-molecular complex $(pp\mu e)^*$ with the proton $p$ and electron $e$. This complex splits apart \cite{kor96} by either direct dissociation
\begin{equation}
    (pp\mu e)^* \rightarrow (\mu p)^* + (pe)
\end{equation}
or by electron emission and subsequent dissociation
\begin{equation}
    (pp\mu e)^* \rightarrow (pp\mu)^* + e \rightarrow  (\mu p)^* + p + e \,.
\end{equation}
The formed muonic hydrogen atom $(\mu p)$ quickly de-excites to the ground state and is thermalized by multiple scatterings on the surrounding hydrogen molecules. Due to the high-pressure the diffusion radius of the muonic hydrogen atom is limited to only about 0.5~mm at the given conditions. The key to increasing the mobility and distance that the muons can travel lies in the small admixture of deuterium that typically amounts to 0.25\% at the chosen pressure of 100~bar. Once the muonic hydrogen atom scatters with deuterium the muon is quickly transferred -- due to the increased binding energy -- forming muonic deuterium $(\mu d)$ in the $1s$ ground state, where $d$ is the deuteron:
\begin{equation}
    (\mu p) +d \rightarrow (\mu d) + p
    \label{eq:transferReaction}
\end{equation}
The rate of this transfer at the present conditions is $\sim 3.5 \times 10^6$/s \cite{and07}. Because of the increased binding energy the formed muonic deuterium atom is accelerated resulting in a kinetic energy of 45~eV \cite{Mul06}. This energy is quickly reduced in subsequent scatterings. However, due to the Ramsauer-Townsend effect \cite{ram21} the transport scattering cross section\footnote{For a definition of the transport cross section see, e.g., Ref. \cite{ada07}.} of $\mu d$ on H$_2$ features a deep minimum at around 4~eV as shown in Fig.~\ref{fig_scattering-XS} \cite{Ada96, Mul06, ada07}. This effect increases the range of the $\mu d$ atom in the gas cell by about a factor 10 over the one of the thermalized $\mu p$ atom allowing some of the muons to reach the microgram target material mounted at the back of the gas cell.

Once the $\mu d$ atom reaches the target material, the tightly-bound and neutral atom easily penetrates the electron cloud and the muon quickly transfers yet again to the target nucleus $Z$ \cite{ger62, pon67}:
\begin{equation}
    (\mu d) + Z \rightarrow (\mu Z)^* + d
\end{equation}
The transfer inside the solid target material is not well understood theoretically and has not been studied in great detail experimentally. Estimates for the transfer rate can be obtained from the calculations of Refs.~\cite{fio76, bra78}. For gold the estimated transfer rate is $\sim 1.6 \times 10^{12}$/s as given by the equation in the conclusions of Ref.~\cite{fio76}. Neglecting any scattering effects and assuming an energy of the $\mu d$ atom of 4~eV, this translates to a characteristic length of 12~nm at which the fraction of remaining $\mu d$ atoms has dropped in gold to $1/e$.

\section{Simulation}\label{sec_simulation}

An efficient measurement requires as many muons as possible to reach the target nuclei after having stopped in the target gas and formed muonic hydrogen atoms. We have conducted Monte Carlo simulations of the transport process to optimize the target geometry and the gas conditions. In order to simulate the drift of $\mu p$ and $\mu d$ atoms through the gas volume, it is necessary to quantify the collisions between the atoms in the $1s$ state and the molecules of the target gas. In addition to the transfer reaction discussed in Equation \eqref{eq:transferReaction}, the following scattering processes take place:
\begin{equation}
\mu a \left(F\right)+ M_2 \rightarrow \mu a \left(F^{\prime}\right)+ M_2
\end{equation}
Here, $a$ denotes a proton or deuteron bound to a negative muon and $M_2$ denotes either one of the molecules H$_2$, D$_2$ or HD. During the scattering process, the hyperfine state $F$ of the muonic atom can be changed when the $\mu a$ atom undergoes a spin-flip reaction in which the atomic nucleus $a$ is exchanged with one of the nuclei of the molecule. At collision energies of~eV and more, the differential cross sections for the molecular scattering processes are approximately equivalent to those for the corresponding nuclear processes, which consider three-body Coulomb interactions between the muonic atom and the molecular nuclei and have been addressed in numerous efforts \cite{Ponomarev1979,Bracci1989,Bracci1990,Bubak1987}. In the experiment, however, the muonic hydrogen atoms can have kinetic energies well below $0.1~$eV for a significant amount of time. In this regime, molecular binding effects, electron screening and spin correlations of specific rotational states can not be neglected any more. Furthermore, excitations and de-excitations of rotational and rotational-vibrational molecular states introduce additional inelastic components to the collisions. For this reason, molecular cross sections need to be used for a realistic simulation of the diffusion process. The method of calculating such molecular cross sections has been described in detail in \cite{Adamczak2006}.

The Monte Carlo simulations are based on the implementation of custom molecular scattering processes in the Geant4 framework \cite{AGOSTINELLI2003250} with use of the G4beamline program \cite{Roberts2007G4beamline}. The implemented scattering processes work with double-differential scattering rates in the laboratory reference frame which have been obtained from the differential cross sections in the center-of-mass provided by \cite{Ada96}. When transferring to the lab frame, the thermal motion of the target molecules is taken into account by averaging over the Maxwell-Boltzmann distribution of the momenta of the gas molecules. 

In addition to the custom molecular scattering processes, the simulations include the physics processes provided by Geant4 for the interaction of the negative muons with materials in the target region. The simulations start with muons from the beam passing through the entrance detector in front of the target cell. We assume the muons in the beam to be normally distributed around the beam center with a standard deviation of 10~mm. The momentum is distributed normally around the mean momentum with a momentum bite of 4\% (full width at half maximum). Muons that enter the gas volume through the window and stop in the gas are substituted with $\mu p$ atoms in the $1s$ state or directly with $\mu d$ atoms (with low probability corresponding to the deuterium concentration). The muonic atoms undergo multiple collisions with molecules including the transfer reaction while drifting through the gas volume. Eventually, the tracks are killed when reaching the target surface or another surface inside the gas cell, corresponding to the transfer towards heavy nuclei. All simulation results presented in this text correspond to a target disk of 15~mm diameter which is placed 15~mm behind the window of the gas cell.

\begin{figure}%
\centering
\includegraphics[width=0.5\textwidth]{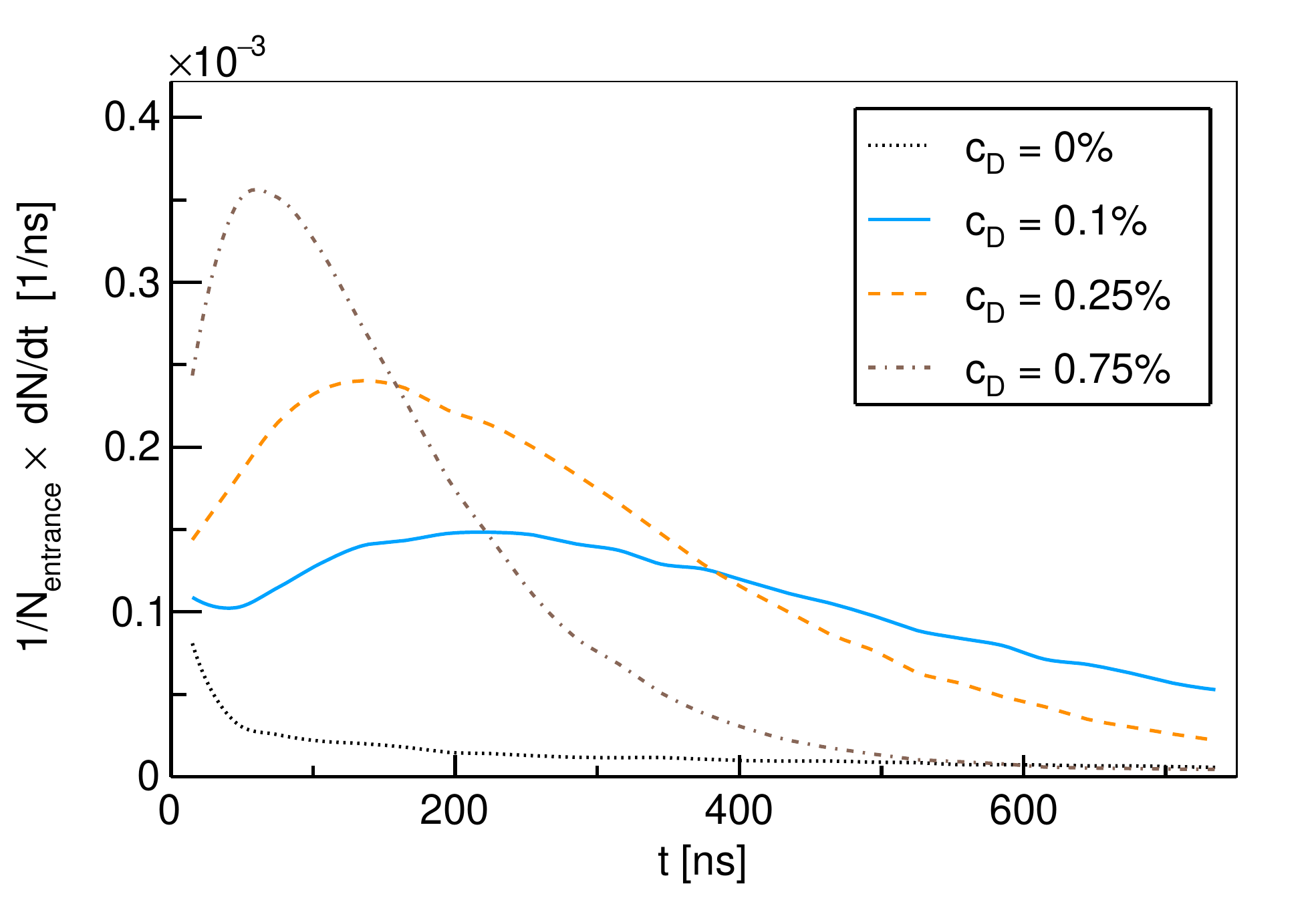}
\caption{Time distributions of $\mu d$ and $\mu p$ atoms reaching the surface of the target disk for various deuterium concentrations $c_D$. The time $t$ corresponds to the time after the incoming muons have triggered the entrance detector. The distributions are normalized by the number of entrance counts.} \label{fig_timeDistribution}
\end{figure}

In reality, the formation process of muonic atoms with the subsequent fast de-excitation cascade is highly complex. Collisions of the large, excited muonic atoms with surrounding molecules influence the kinetic energy distribution of the atoms after reaching the $1s$ state. Based on the standard cascade model \cite{Leon1962}, different theoretical models of the cascade process have been developed \cite{Borie1980,Markushin1984,Jensen2002,Faifman2008,Popov2012,Popov:2017PRA,Popov:2021PRA}. Their results for the kinetic energy of muonic atoms after the cascade, however, show strong variations. In accordance with experimental findings \cite{Abb97,poh10}, the most recent calculations \cite{Popov:2017PRA,Popov:2021PRA} find that a part of the atoms ends up thermalized after the cascade while other atoms cover the whole energy range up to several tens of eV, with few atoms even reaching energies of 100 to 200 eV.
For the results of the transport simulations described here, no significant effect of the initial energy distribution could be observed. Due to the high density of around 10\% of liquid hydrogen, the $\mu p$ atoms at a kinetic energy of 100~eV have mean free paths of only a few \si{\micro\metre}, which makes them slow down after minimal distances. For the simulation results presented in this text, the initial kinetic energy of the muonic atoms in the $1s$ state was assumed to be distributed according to a Maxwell-Boltzmann distribution with a mean energy of 1~eV.

\begin{figure}%
\centering
\includegraphics[width=0.5\textwidth]{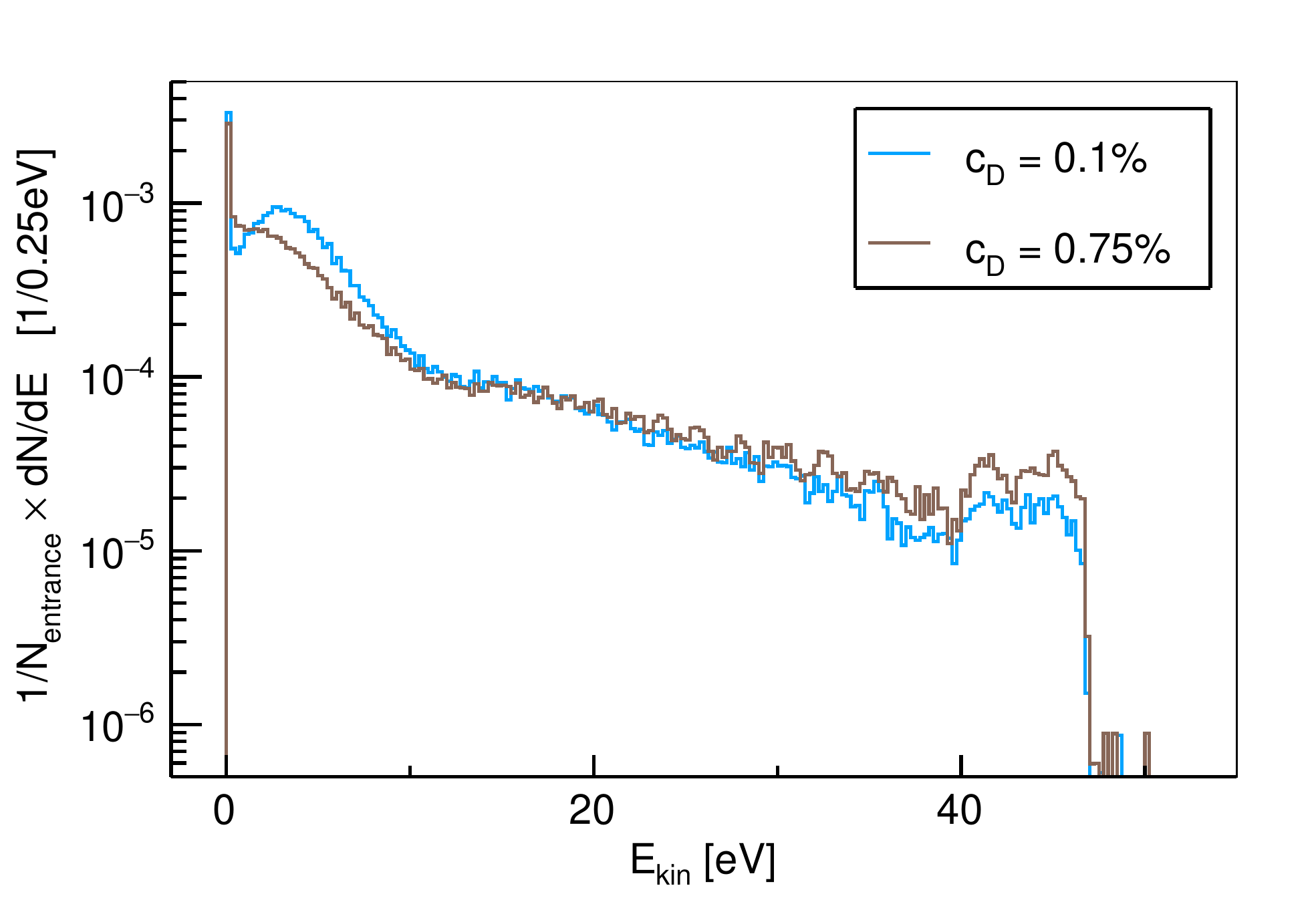}
\caption{Energy distributions of $\mu d$ and $\mu p$ atoms reaching the surface of the target disk for two different deuterium concentrations $c_D$. Only atoms that reach within the first $1~\mu$s after the entrance detector has been triggered are considered here. The distributions are normalized by the number of counts in the entrance detector.} \label{fig_energyDistribution}
\end{figure}

Time distributions of muonic hydrogen atoms reaching the target disk are presented in Fig.~\ref{fig_timeDistribution} for various deuterium concentrations in the hydrogen gas. The distributions were obtained for a muon beam with a mean momentum of $28.0~$MeV/c and have been normalized to the number of entrance counts. The distribution for pure hydrogen ($c_{\mathrm{D}}=0\%$) shows a prompt peak stemming from $\mu p$ atoms that reach the target disk directly after formation. The other distributions show significant peaks at later time which are associated with $\mu d$ atoms reaching the cell wall after the isotopic exchange of the muon. Increasing deuterium concentration raises the rate of the isotopic muon exchange from $\mu p$ to $\mu d$, which enables the $\mu d$ atoms to reach the target disk faster. On the other hand, a higher concentration of deuterium in the mixture increases the probability for $\mu d$ atoms to scatter with deuterium molecules. The scattering cross sections for this process exhibit no Ramsauer-Townsend minima (see Fig.~\ref{fig_scattering-XS}) and can therefore limit the free path of the $\mu d$ atoms. The optimal deuterium concentration which balances these concurrent effects typically lies between 0.1 and 0.25\% depending on the exact target and beam conditions. Energy distributions of the muonic atoms at the moment they reach the target disk are shown in Fig.~\ref{fig_energyDistribution} for two different deuterium concentrations. As can be seen from the distributions, most $\mu d$ atoms have lost the energy gained in the isotopic exchange reaction when reaching the target. A peak of the energy distribution around the Ramsauer-Townsend minimum at $\sim 4~$eV is visible for the lower deuterium concentration ($c_{\mathrm{D}}=0.1\%$). For the higher concentration, this peak has been washed out due to the increased scattering on deuterium molecules. 

\begin{figure}%
\centering
\includegraphics[width=0.5\textwidth]{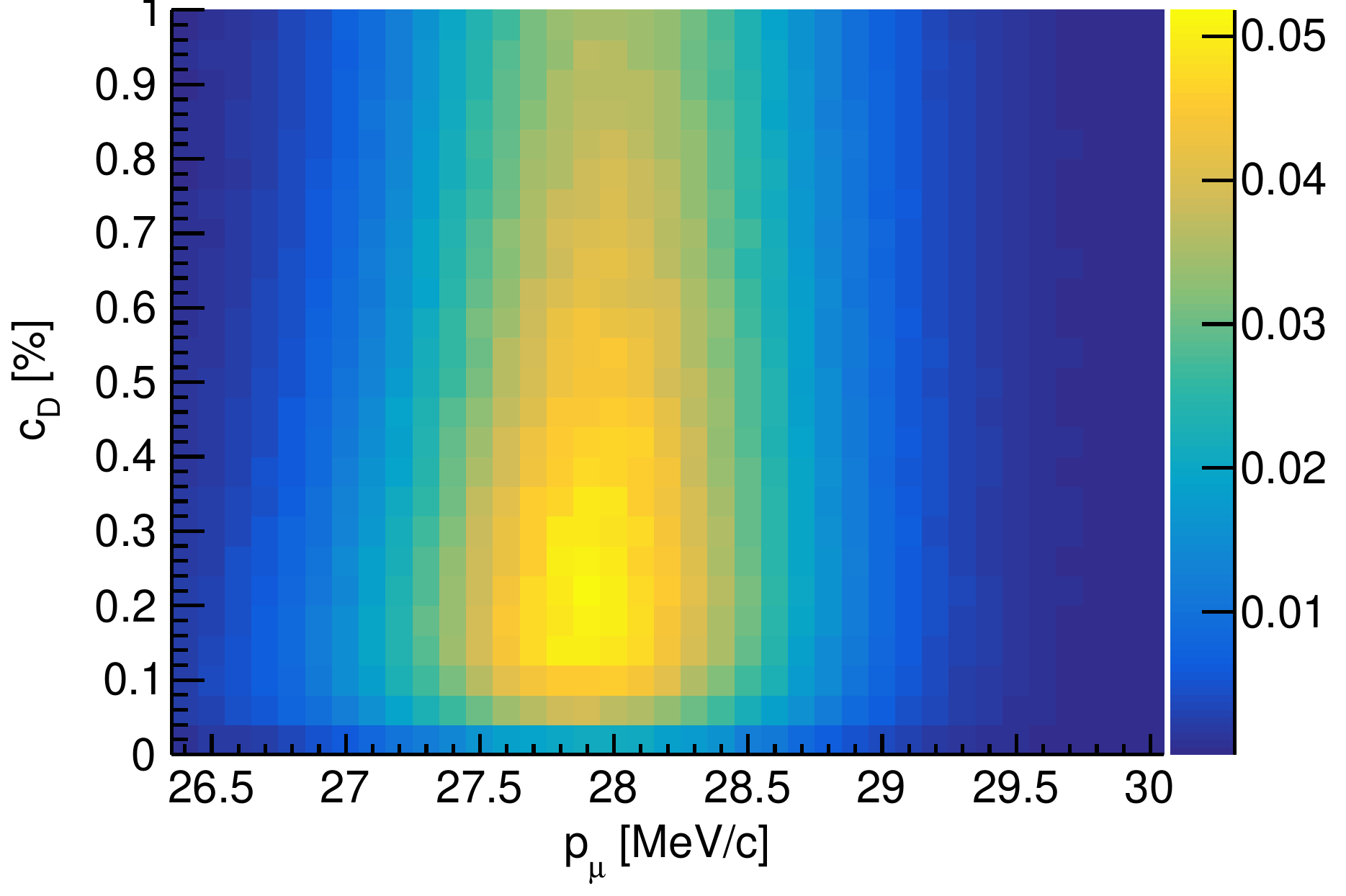}
\caption{Parameter scan for the momentum of the muon beam $p_{\mu}$ and the deuterium concentration $c_D$. The figure of merit is the probability for muons which are detected in the entrance detector to eventually be transferred to the target disk within maximally 1~\si{\micro\second}. In the simulations, a gaussian muon beam momentum distribution with 4\% full width at half maximum was assumed.} \label{fig_p_cD_Scan_2019}
\end{figure}

Figure~\ref{fig_p_cD_Scan_2019} shows a parameter scan performed with the transport simulations. Here, the efficiency of the transport process is defined as the number of muonic atoms reaching the target disk normalized by the number of muons passing through the entrance detector. For this, only muonic atoms are considered that reach the target disk within 1000~ns after entering the target cell. Note, that the transfer to the heavy nuclei in the target is not considered in the simulations. In reality and depending on the thickness of the target, only a fraction of the muonic hydrogen atoms reaching the target surface transfer the muon to the actual target nuclei, while the rest of the muons are transferred to the target backing material. For the scan, the efficiency has been determined for various beam momenta and deuterium concentrations. For this geometry, a maximal efficiency is reached with an average momentum of 27.8~MeV/c -- at which point a significant fraction of the muons stop directly in the target backing material -- and a deuterium concentration of 0.15\%. Comparisons of simulation results with experimental data will be presented in Sec.~\ref{sec:results} of this text.

\section{Apparatus}\label{sec_apparatus}

\begin{figure}%
\centering
\includegraphics[width=0.5\textwidth]{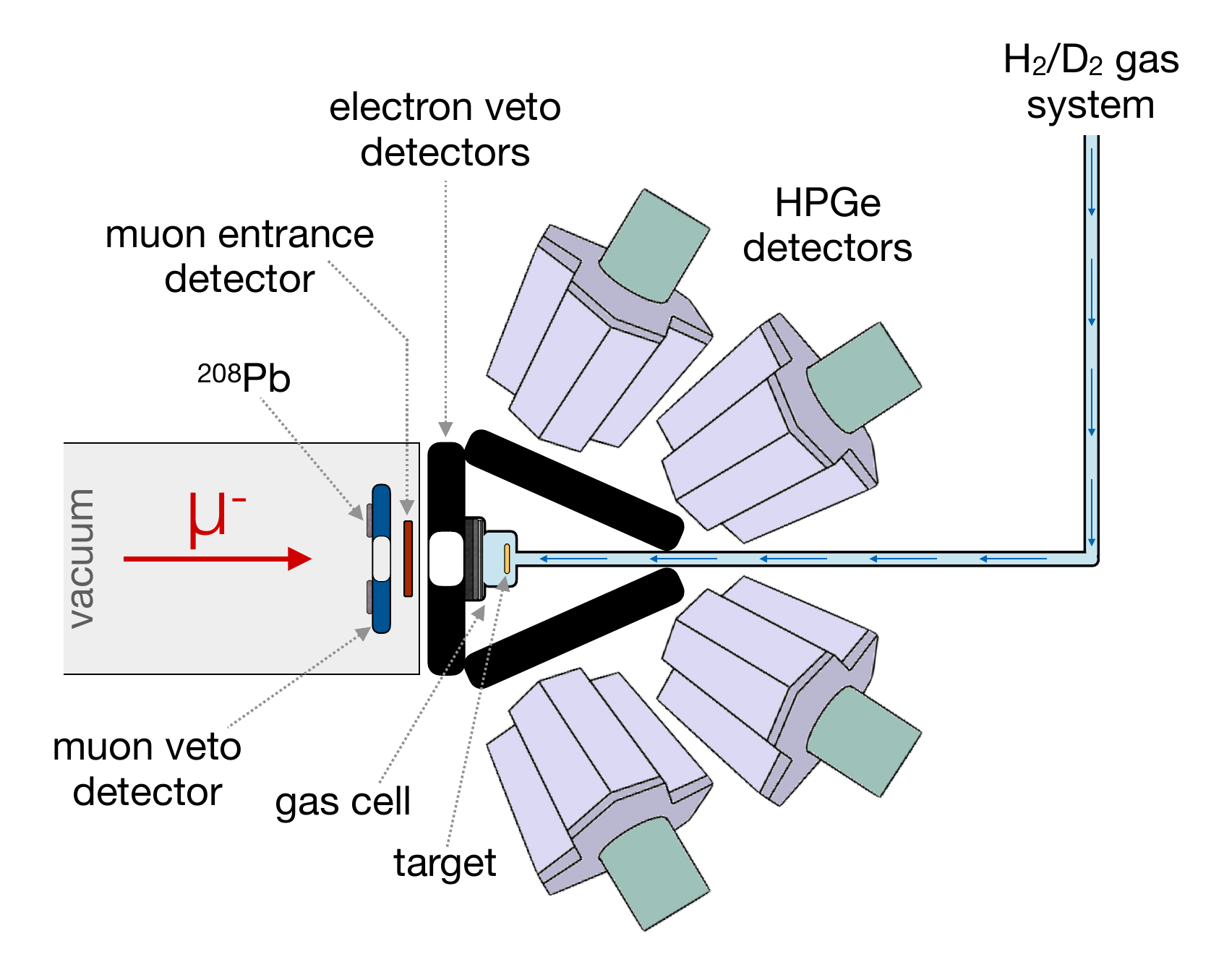}
\caption{Sketch of the muX apparatus with its different components labelled. For more details see the dedicated sections.} \label{fig_muX-sketch}
\end{figure}

Figure \ref{fig_muX-sketch} shows an overview of the apparatus and its different components, which has been discussed in parts also in Refs. \cite{Ada18, ska19, Knecht:2020npz, Wauters:2021cze}. The different components and the layout were developed in order to enable the method described in Section~\ref{sec_method} with the central gas cell featuring a geometry and dimensions optimized through the
simulation described in Section~\ref{sec_simulation}. The following subsections describe the different elements in detail.

\subsection{Muon beamline}
The measurements were performed at the $\pi$E1 beamline in the experimental hall of PSI's high-intensity proton accelerator HIPA \cite{Grillenberger:2021kyv}. A static $\mathbf{E} \times \mathbf{B}$ separator installed at the beginning of the experimental area allowed to reduce the electron contamination in the beam to negligible levels thereby providing a very pure negative muon beam at the chosen momentum of around 28~MeV/c. Using a slit system mounted in the beamline at a point of large dispersion allowed to adjust the momentum acceptance of the channel to around 3-4\% full width at half maximum.

The apparatus was installed about 1.3~m downstream of the last quadrupole triplet of the $\pi$E1 beamline. At this position the beamspot had a size of about 10~mm (gaussian sigma) in horizontal and vertical direction. Typical rates on the muon entrance scintillator (see Sec.~\ref{sec_scintillators}) located behind the beam veto scintillator (see Sec.~\ref{sec_scintillators}) with an opening of 18~mm were around 20~kHz at a proton beam intensity of 2~mA.

\subsection{Beam veto, muon entrance and decay-electron veto scintillators}\label{sec_scintillators}

\begin{figure*}%
\centering
\begin{tabular}{c c c}
\includegraphics[width=0.25\textwidth]{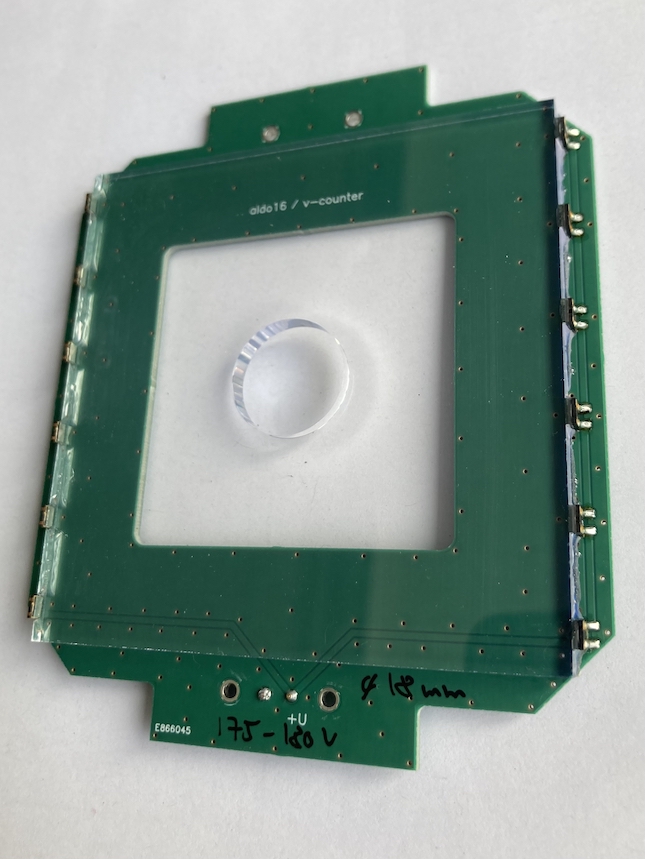} & 
\includegraphics[width=0.25\textwidth]{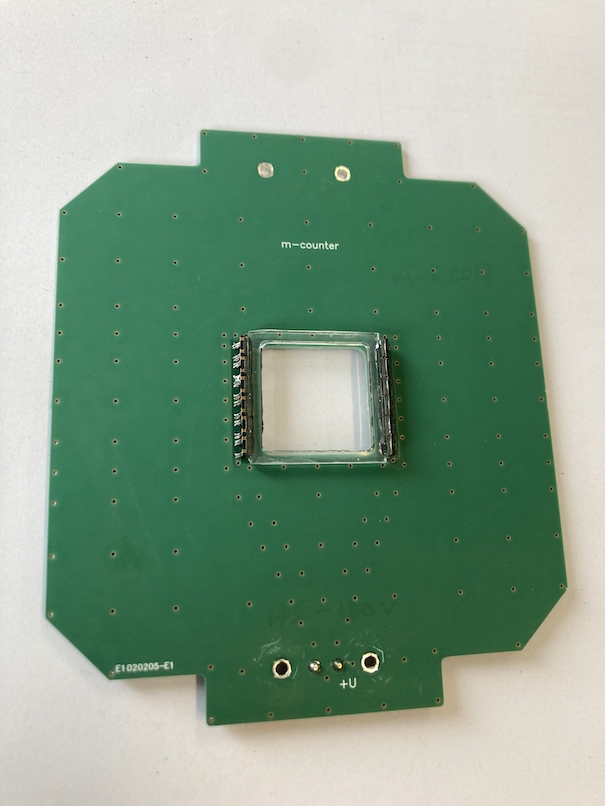} & 
\includegraphics[width=0.25\textwidth]{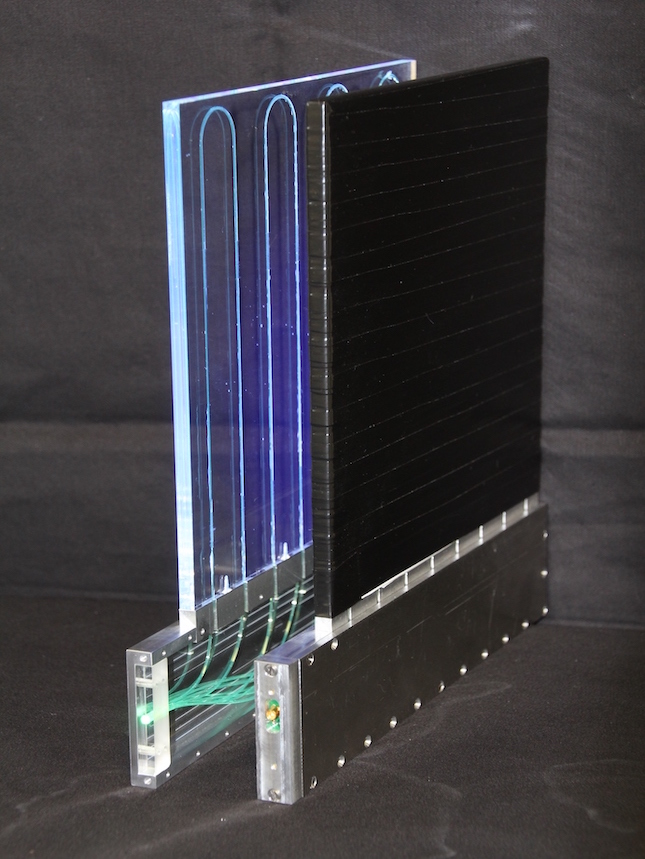} \\
(a) & (b) & (c)
\end{tabular}
\caption{Plastic scintillator detectors used in the muX apparatus: a) beam veto scintillator, b) muon entrance scintillator, and c) decay-electron veto scintillators. For more details see text.} \label{fig_scintillators}
\end{figure*}

As shown in Fig.~\ref{fig_muX-sketch}, the muX apparatus employs three different plastic scintillator detectors. They are presented in Fig.~\ref{fig_scintillators}. The first detector in beam direction is the beam veto scintillator (Fig.~\ref{fig_scintillators} a). This is based on a $80\times 80$~mm$^2$ BC-404 scintillator plate with a thickness of 4~mm and a central hole with a diameter of 18~mm. It is read out on two sides by a total of 12 $3\times 3$~mm$^2$ silicon photomultipliers (model ASD-NUV3S-P by AdvanSiD \cite{advansid}) mounted on a PCB. This detector serves as an active collimator for the muon beam and vetoes against background coming down the beamline outside of the acceptance of the entrance detector.

The beam veto detector is followed by the muon entrance detector (Fig.~\ref{fig_scintillators} b). The detector is based on a $20\times 20$~mm$^2$ BC-400 scintillator foil with a thickness of 200~\si{\micro\metre}. The scintillator is glued with optical cement on a UV-transparent plexiglas frame, which in turn is coupled to 12 $3\times 3$~mm$^2$ silicon photomultipliers as for the beam veto detector. This detector serves to provide a precise timing for the arrival of a muon in the high-pressure gas cell and to distinguish a muon from some remaining electron background in the beam through the amount of deposited energy.

The last scintillator detectors are the decay-electron veto detectors shown in Fig.~\ref{fig_scintillators}c that surround the gas cell. These are large  $180\times 180$~mm$^2$ EJ204 scintillator plates with a thickness of 5~mm. The light from the scintillator is collected through wavelength shifting fibers that are embedded into grooves in the scintillator plate. They guide the light to a single $3\times 3$~mm$^2$ silicon photomultiplier (model S13360-3050CS by Hamamatsu \cite{hamamatsu}). One of the detectors features a 35~mm diameter central hole such that it can be mounted around the tube through which the muons enter the gas cell (see Fig.~\ref{fig_muX-sketch}). These decay-electron veto detectors serve to efficiently reduce the background generated in the germanium detectors (directly or indirectly) coming from muon Michel decays generating electrons with energies up to 53~MeV.

\subsection{High-pressure gas cell}\label{sec_gas_cell}

\begin{figure}%
\centering
\includegraphics[width=0.45\textwidth]{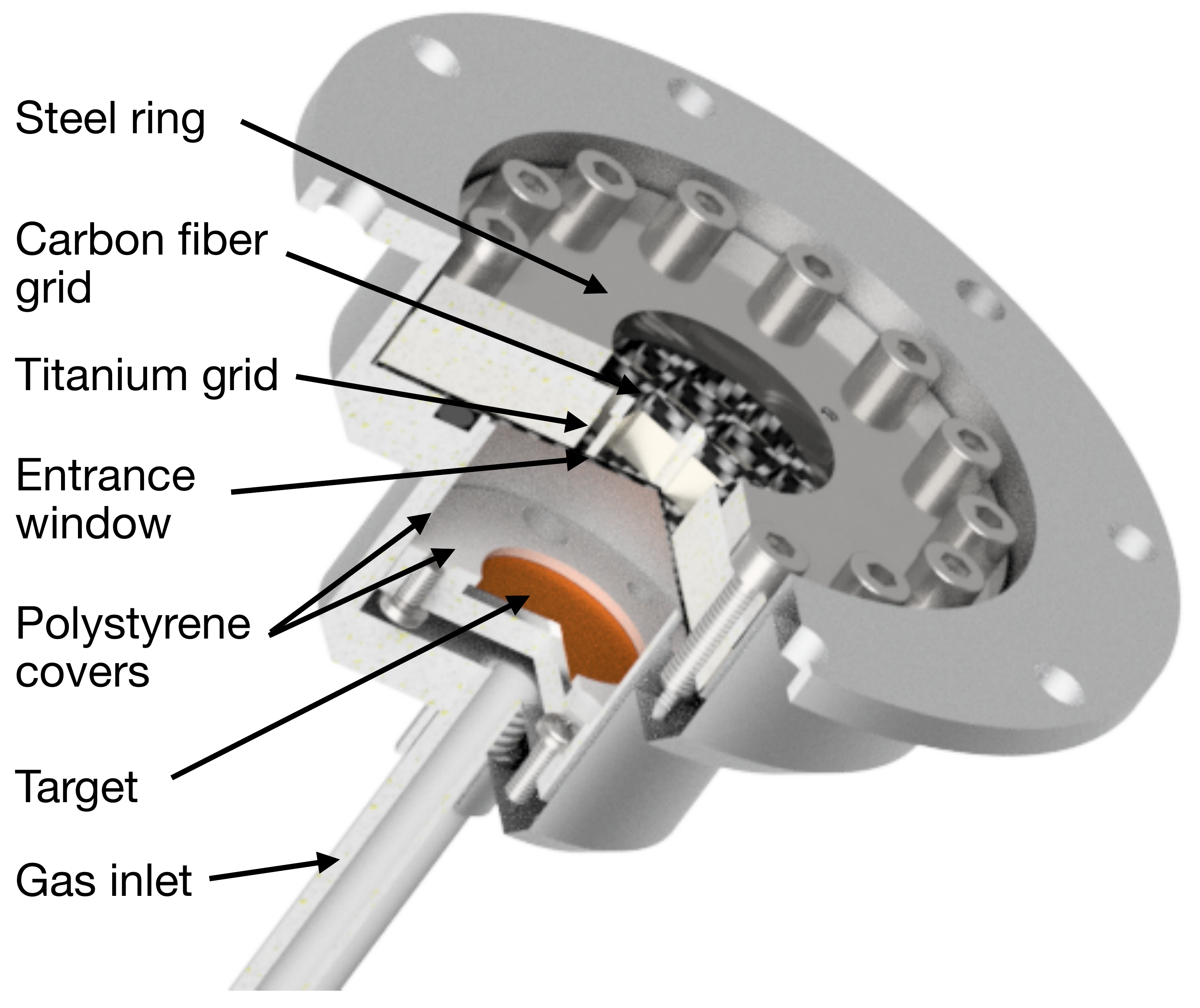}
\caption{Rendered CAD image of the high-pressure gas cell with a cutout in order to show the placement of the target and the arrangement of the entrance window and supporting grids. For reference the volume between the target and entrance window has a length of 14~mm and a diameter of 30~mm. See text for more information on the different parts.} \label{fig_CellAssembly}
\end{figure}

Figure~\ref{fig_CellAssembly} shows an overview of the high-pressure gas cell. The main body of the cell is made of aluminum. An aluminum gas line screws into the main body from the back (feeding in the gas just behind the target) and is sealed using glue (Model Stycast 2850 FT). The cell is closed towards the muon beamline by a 600~\si{\micro\metre} thick carbon fiber-reinforced polymer window that allows the muons to pass through. In order to withstand the high pressure, the window is supported by a combination of a first 4~mm thick titanium grid and a second 2~mm carbon fiber-reinforced polymer grid mounted on top. The 2~mm carbon fiber-reinforced polymer grid faces the muon beamline and greatly reduces the background from muons stopping in titanium. The grids have an open area of 74\%. During measurements a transmission through this grid and window arrangement of around 60\% was observed (with a variation of around $\pm$5\% over the various experimental campaigns depending on beam setup) and thus smaller than the open area due to the divergence of the beam and stops in the window. The grids and window are pressed down by a 3~mm thick stainless steel ring with an open diameter of 20~mm. The overall gas seal is achieved by an O-ring. The cell was pressure tested up to 350~bar at which point the screws securing the stainless steel ring and titanium and carbon fiber grids are ripped out of the aluminum body. 

A polystyrene tube (not shown in Fig.~\ref{fig_CellAssembly}) extends from the stainless steel ring to the entrance detector ensuring that muons can only stop in low-$Z$ material on their passage from the entrance detector to the gas cell. Also the inside of the gas cell is lined with a polystyrene tube removing any potential background from muonic hydrogen or deuterium atoms reaching the aluminum walls of the pressure cell.

The inner space of the gas cell has a diameter of 30~mm and a length of 20~mm. At the end of this space the target in its target holder is mounted reducing the distance between the entrance window and target surface to 14~mm, which is the available length for the muons to stop in the gas.

\subsection{Hydrogen/deuterium gas system}\label{sec_gas_system}
A dedicated gas handling system was built that allows the controlled operation and mixing of gases from three different sources and the filling of the gas cell described above through an approximately 3~m long stainless steel tube with 4~mm inner diameter. It uses various absolute pressure gauges that also monitor the pressure inside the gas cell during normal operation. Before filling the gas cell with the hydrogen/deuterium mixture, the cell is flushed several times with hydrogen in order to reduce residual air contamination and impurities in the gas as much as possible.

In 2017/2018, the gas mixture was prepared by first filling the tube and gas cell with the desired pressure of deuterium and in a next step filling hydrogen up to a pressure of 100~bar. By comparing the measurements of the transfer reactions with the simulation (and later also by performing Raman spectroscopy on the deuterium/hydrogen gas mixtures), it was realized that this procedure leads to much higher and somewhat unreproducible deuterium concentrations as the filling with hydrogen compresses all the deuterium left in the filling tube into the gas cell. Due to the high pressures the diffusion times leading to an equal mixture are very high and thus no equilibrium is reached before the start of the measurements \cite{nub18}. From 2019 onward, we thus switched to the use of pre-mixed gases obtained commercially.

\subsection{Target}\label{sec_target}

\begin{figure}%
\centering
\includegraphics[width=0.3\textwidth]{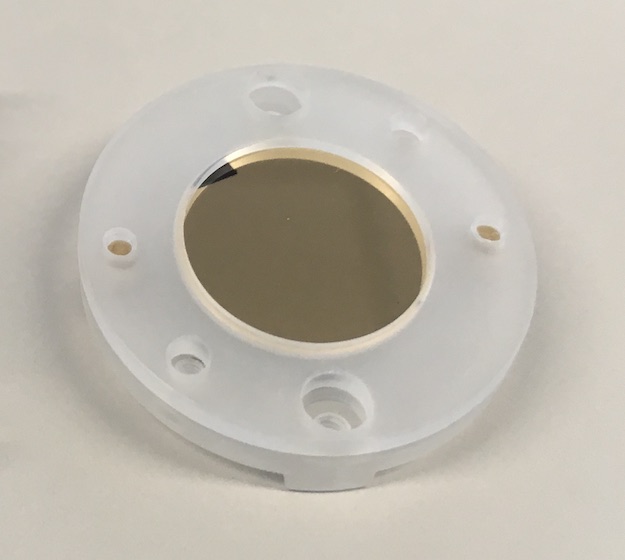}
\caption{Typical gold target used during the measurements described in this paper mounted in its polystyrene holder.} \label{fig_gold_target_holder}
\end{figure}

The backing disks for the target materials are made of glassy carbon (Sigradur from HTW \cite{htw}) with a diameter of 16~mm and a thickness of 1~mm. The gold targets employed for the measurements described here were manufactured by chemical vapor deposition using the in-house facilities available at PSI.

The glassy carbon disks are mounted in a holder made of polystyrene, which as described above is fastened at the back of the gas cell using PEEK screws. The holder reduces the open area of the target to a diameter of 15 mm. In order to avoid any potential contamination of the gas in the cell the screws were vented through small holes and generally only procedures applicable for high-vacuum environments were employed during the design, construction and handling of the high-pressure gas cell.

\subsection{High-purity germanium detector array}
\begin{figure}%
\centering
\includegraphics[width=0.5\textwidth]{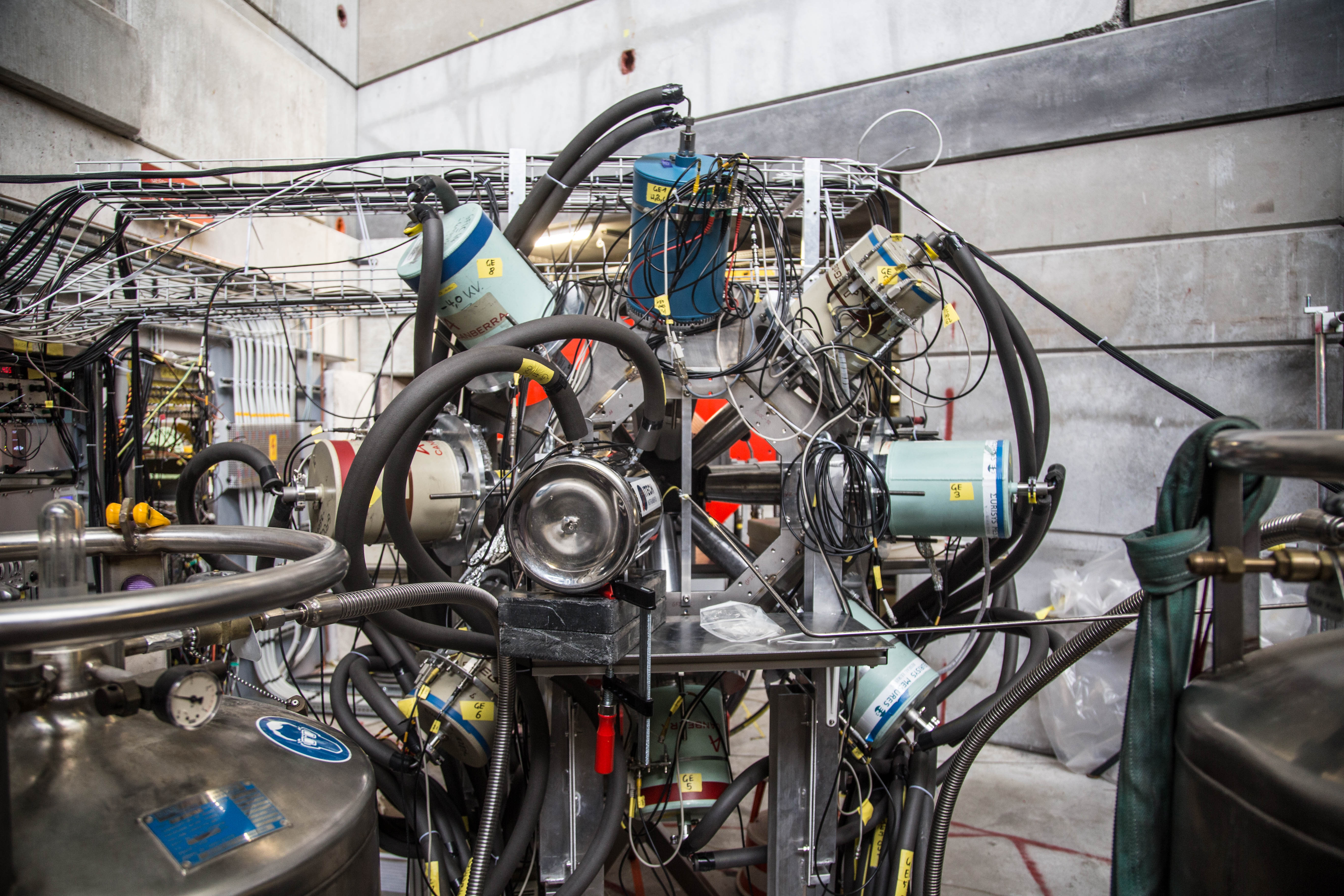}
\caption{Germanium detector array used during the 2017/2018 experimental campaigns.} \label{fig_setup_2017-2018}
\end{figure}
\begin{figure}%
\centering
\includegraphics[width=0.5\textwidth]{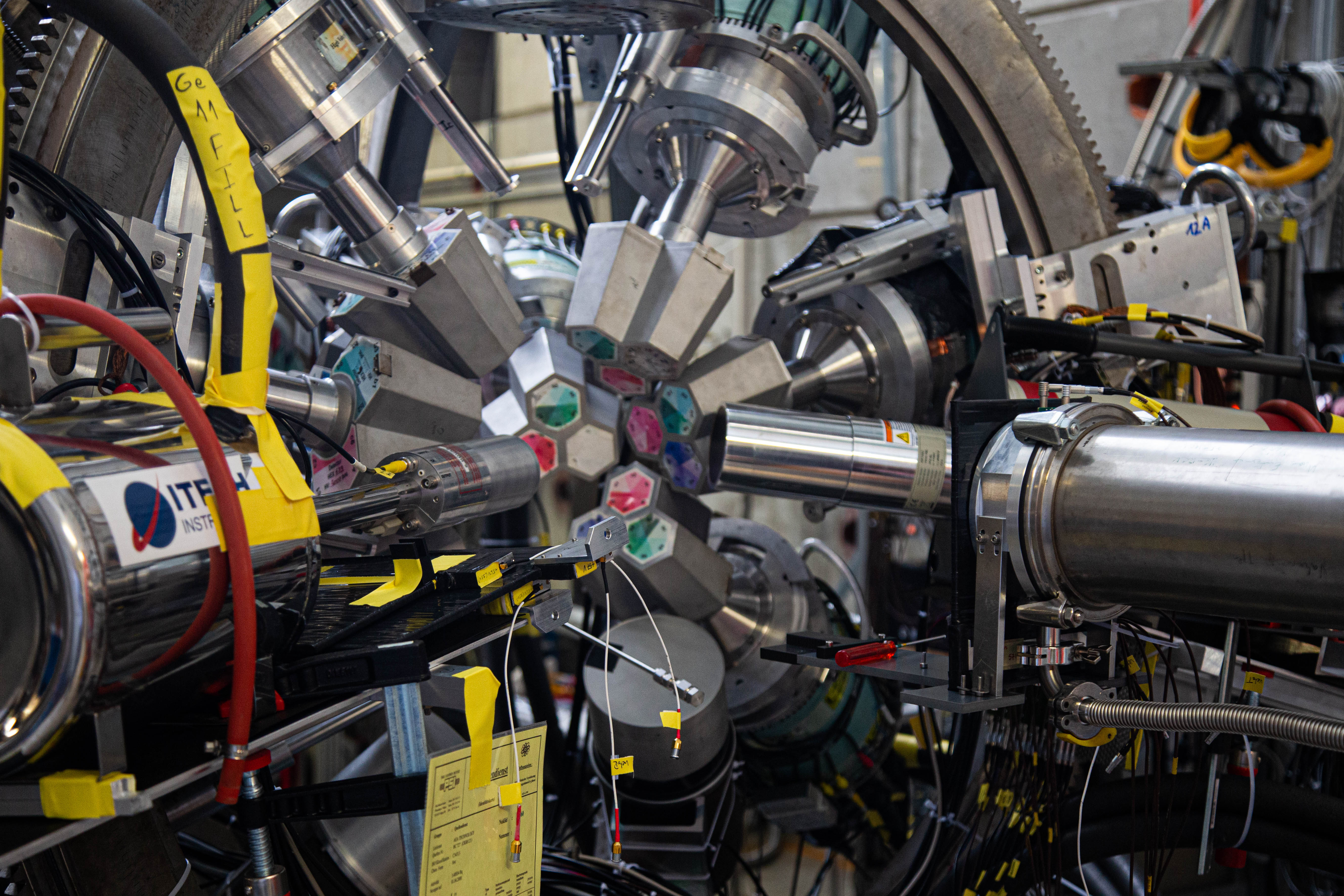}
\caption{Miniball germanium detector array mounted in its custom, highly adjustable frame in the $\pi$E1 area. The photograph shows the array moved backwards on a dedicated rail system in order to gain easy access to the muon entrance detector and gas cell.} \label{fig_setup_2019}
\end{figure}
The germanium detector setups evolved over the course of the different experimental campaigns. In 2017 and 2018 (see Fig.~\ref{fig_setup_2017-2018}), the array was mounted in a custom frame on rails and consisted of seven single-crystal coaxial detectors with $\sim$60\% efficiency from the IN2P3/STFC French/UK Ge loan pool \cite{loan_pool}, two single-crystal coaxial Ge detector with 75\% and 70\% efficiency (an additional 15\% detector was mounted in 2018), a Miniball cluster consisting of three individual crystals of around 60\% each \cite{war13} and a planar detector optimized for low-energy gamma rays. The frame allowed a modest flexibility in positioning the detectors at distances of around 10 to 15~cm away from the target. 

In 2019 due to the long shutdown at CERN, the full Miniball detector array \cite{war13} was brought from ISOLDE to PSI and mounted for an extended period in the $\pi$E1 area. The full array consisted of 8 clusters of three crystals with efficiencies of around 60\% each. The array mounted in its original frame at PSI is shown in Fig.~\ref{fig_setup_2019}. The array was supplemented by the 70\% coaxial and planar detector already used in 2017 and 2018. Due to the larger number, the detectors could only be mounted at a distance of around 15~cm to the target.

With these configurations, we reached about 1.8\% and 2.2\% full energy detection efficiency at 1332~keV with the full arrays for 2017/2018 and 2019, respectively.

\subsection{Calibration} \label{sec_calibration}
While during dedicated calibration runs a large variety of standard gamma calibration sources were used, a scheme was implemented to provide continuous calibration lines during normal muon beam operation. As shown in Fig.~\ref{fig_muX-sketch}, a 100~\si{\micro\metre} thick layer of $^{208}$Pb was mounted behind a 600~\si{\micro\metre} thick carbon fiber-reinforced polymer sheet (the same as the entrance window of the gas cell) -- both with a central hole of 18~mm -- and installed in front of the beam veto detector with the carbon fiber-reinforced polymer sheet facing the beam. This allowed a fraction of the muon beam (those outside the 18~mm acceptance of the beam veto scintillator) to stop in $^{208}$Pb and produce a continuous source of well-known muonic x rays reaching up to high energies \cite{Ber88} for calibration. In addition, a weak 2.6~kBq $^{60}$Co source was placed close to the target that provided additional continuous calibration lines and could also be used for an automated analysis and check of the full detector array by monitoring for each detector the properties of the measured $^{60}$Co lines.

\section{Results}
\label{sec:results}

\subsection{Optimization of transfer process}

The transfer process inside the gas cell is typically optimized through a series of measurements employing a 50~nm thick gold target prepared as described in Sec.~\ref{sec_target}. The 50~nm thickness was chosen under the assumption that such a layer would lead to the full transfer of all muons from the muonic hydrogen and deuterium atoms to gold as they penetrate into the layer. The total target mass amounts to a bit over 200~\si{\micro\gram}, which allows for swift measurements while at the same time only having a negligible amount of direct muon stops inside the gold layer. The optimization procedure basically follows the parameter space outlined in Fig.~\ref{fig_p_cD_Scan_2019} by scanning the deuterium concentration and  muon beam momentum and counting the amount of emitted muonic $2p-1s$ gold x rays at 5591.7 and 5760.8~keV (the higher energy transition is distributed across two lines due to hyperfine splitting) \cite{Ack66,Fri95}.

The observed total counts of $2p-1s$ gold x rays $N_\gamma$ is used to estimate the efficiency $\epsilon_T$ of the full transfer process by relating them to the number of detected muons by the entrance detector $N_\mu$:
\begin{equation}
\epsilon_T = \frac{N_\gamma}{N_\mu} \frac{1}{I \epsilon_{det} \zeta},
\end{equation}
with $I$ the absolute yield per muon of the $2p-1s$ line, $\epsilon_{det}$ the detection efficiency of the full germanium detector array at this energy and $\zeta$ the transmission through the grids and the entrance window of the gas cell (see Sec.~\ref{sec_gas_cell}). 

The yields of the $2p-1s$ transition after transfer is distinctly different and about a factor 2 lower compared to direct capture as was, e.g, measured for several low-$Z$ elements \cite{Ber78, Jac88, Mul93, Jac95}. This can be understood by the fact that the initial distribution across the angular momentum states is shifted to much lower values \cite{Haf77}, which boosts the yields of higher $np-1s$ transitions compared to the direct capture. As no trustworthy measurements of the $2p-1s$ yield after transfer for elements higher than argon exist, it was estimated for gold after transfer to be $65.7\pm 7.1$\%, and thus significantly higher than seen in the measurements at lower atomic numbers, by extrapolating a series of transfer measurements in a hydrogen gas cell with small admixtures of argon, krypton and xenon. A dedicated publication on these measurements is in preparation \cite{Ska21}. 

\begin{figure}%
\centering
\begin{tabular}{c}
\includegraphics[width=0.45\textwidth]{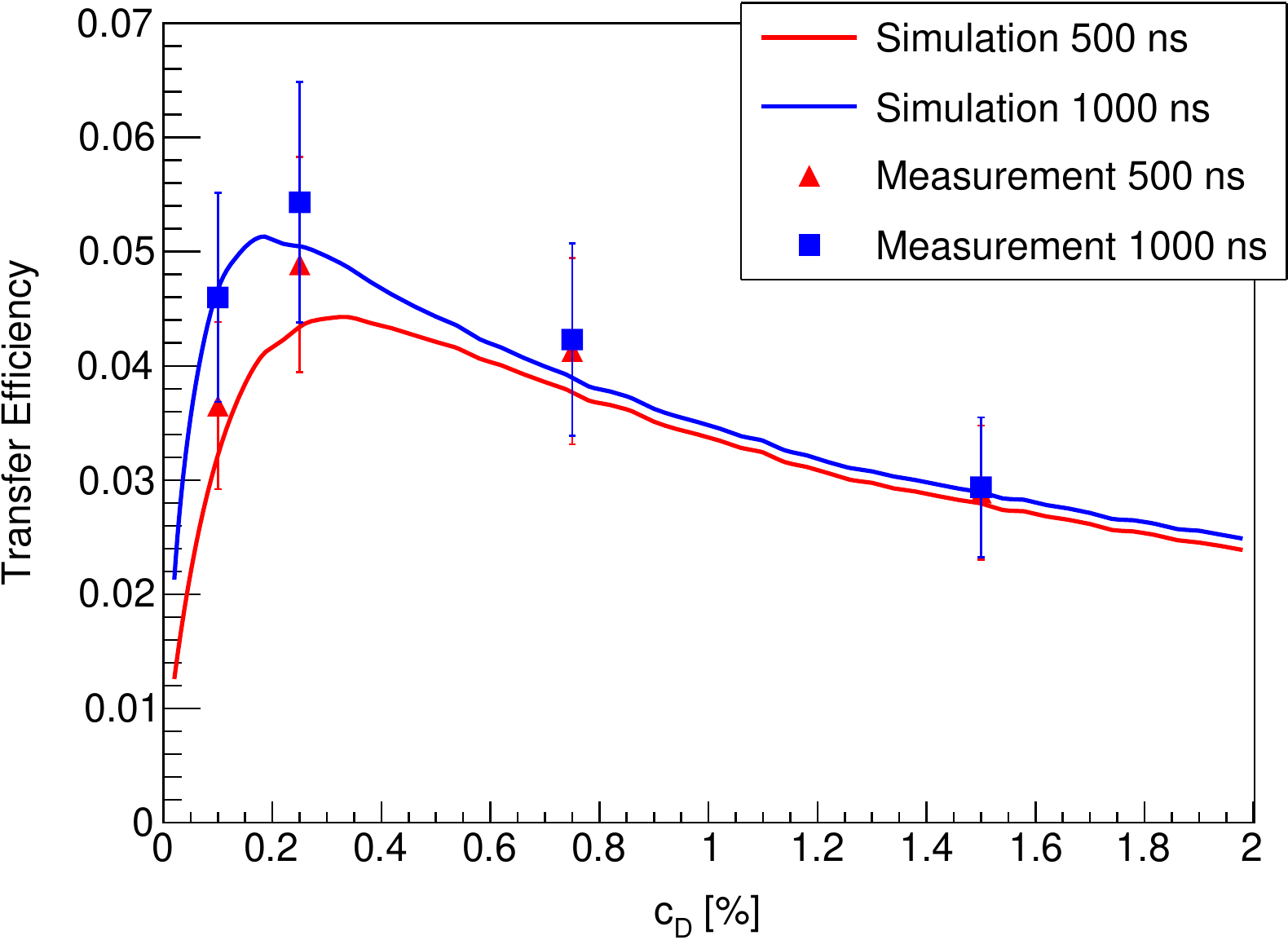} \\
(a) \\
\includegraphics[width=0.45\textwidth]{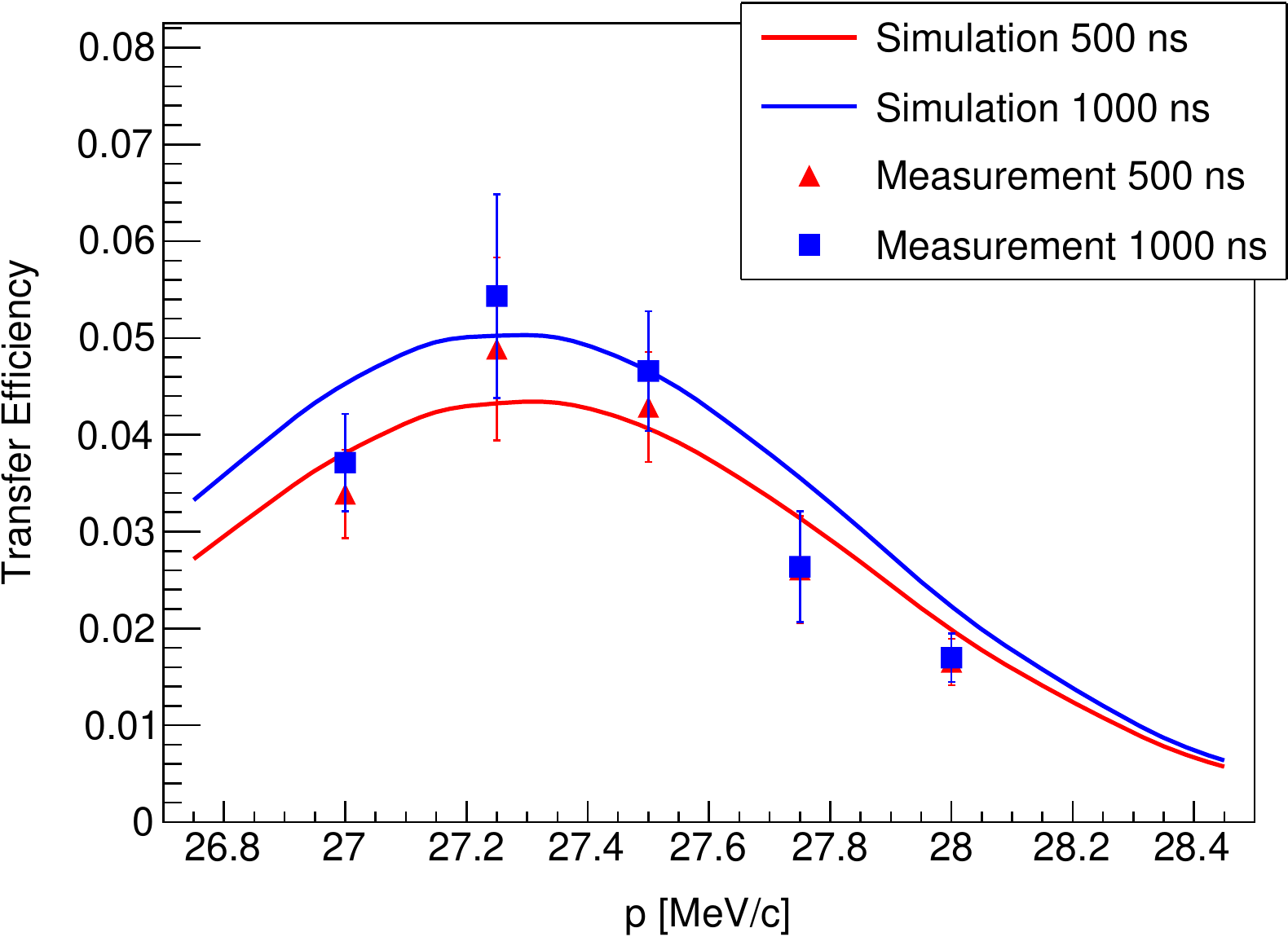} \\
(b)
\end{tabular}
\caption{(a) Transfer efficiency as a function of deuterium concentration $c_D$ at a muon beam momentum of 27.25~MeV/c. (b) Transfer efficiency as a function of muon beam momentum $p$ at a deuterium concentration of 0.25\%. The two time windows correspond to the time after a muon entered the gas cell over which the recorded muonic gold x rays are counted to calculate the transfer efficiency. Also shown are the simulation results from Fig.~\ref{fig_p_cD_Scan_2019} that demonstrate excellent agreement. In the comparisons here, the momentum of the simulated results was shifted by $-0.65$~MeV/c in order to compensate for the not precisely known material budget traversed by the muon beam and to obtain the maximum efficiency at the same momentum as in the measurement.}
\label{fig_scan_cd_p}
\end{figure}

\begin{figure*}%
\centering
\begin{tabular}{c c}
\includegraphics[width=0.4\textwidth]{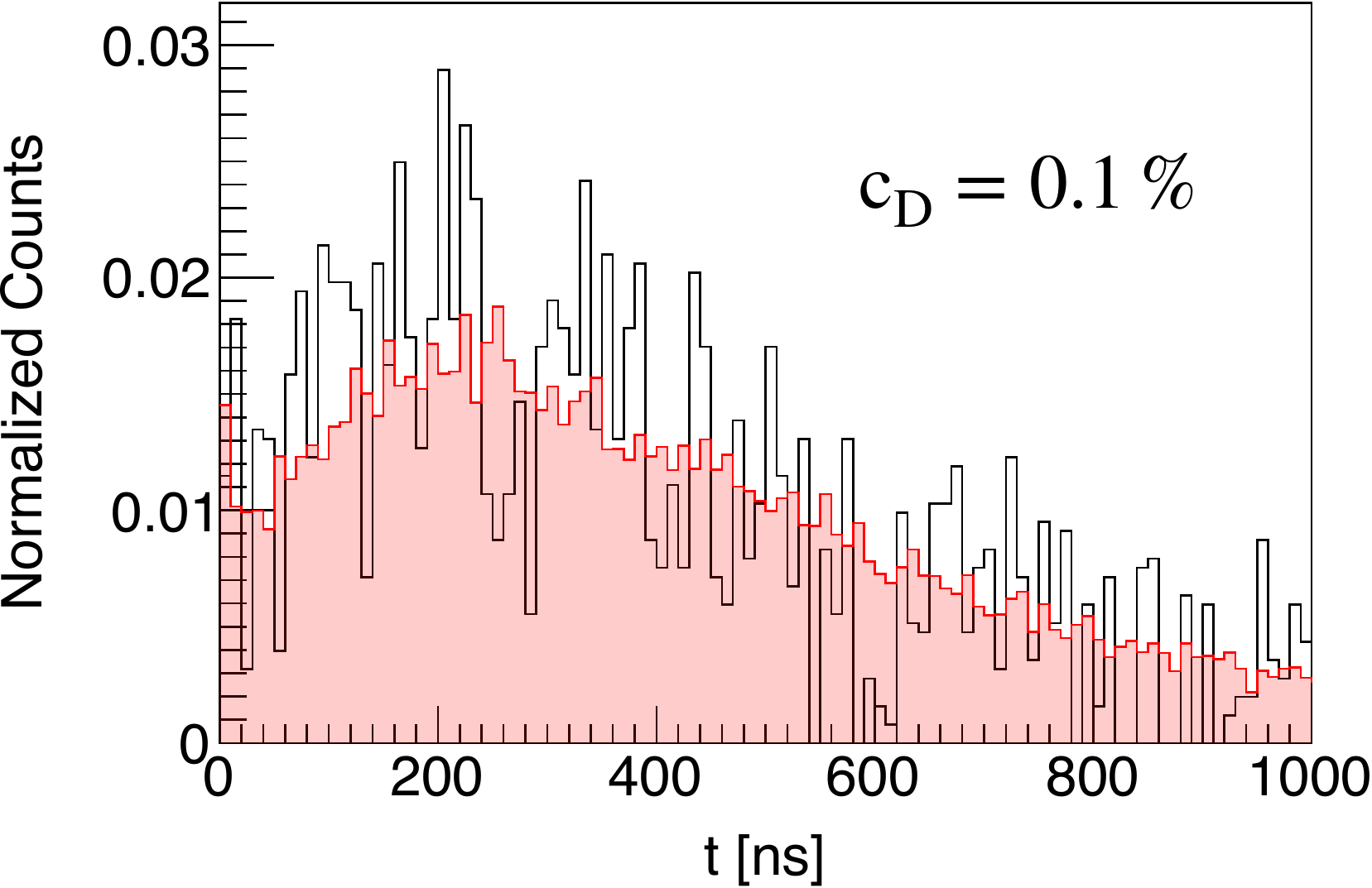} & 
\includegraphics[width=0.4\textwidth]{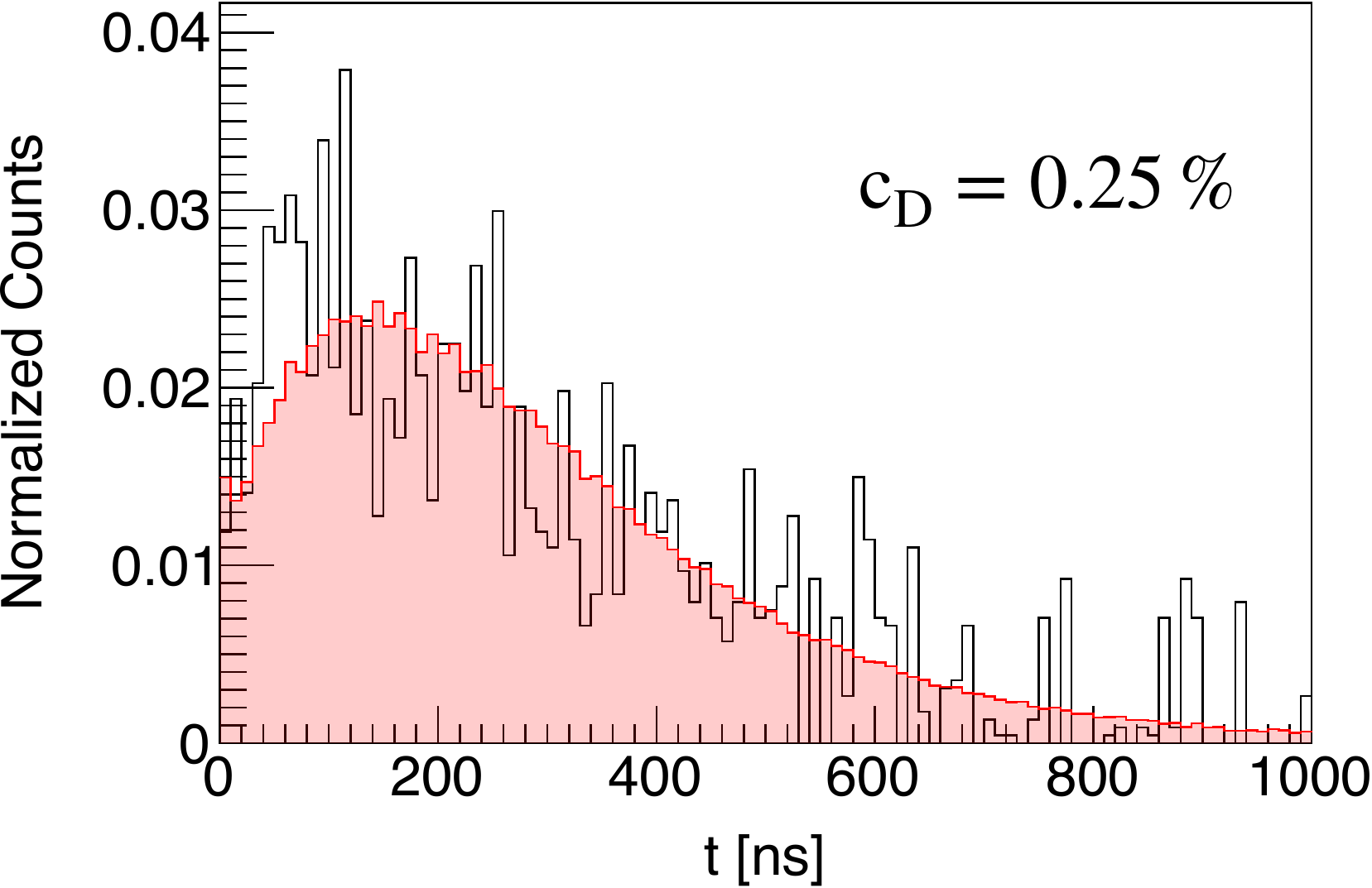} \\ 
(a) & (b) \\
\vspace{0.1cm} \\
\includegraphics[width=0.4\textwidth]{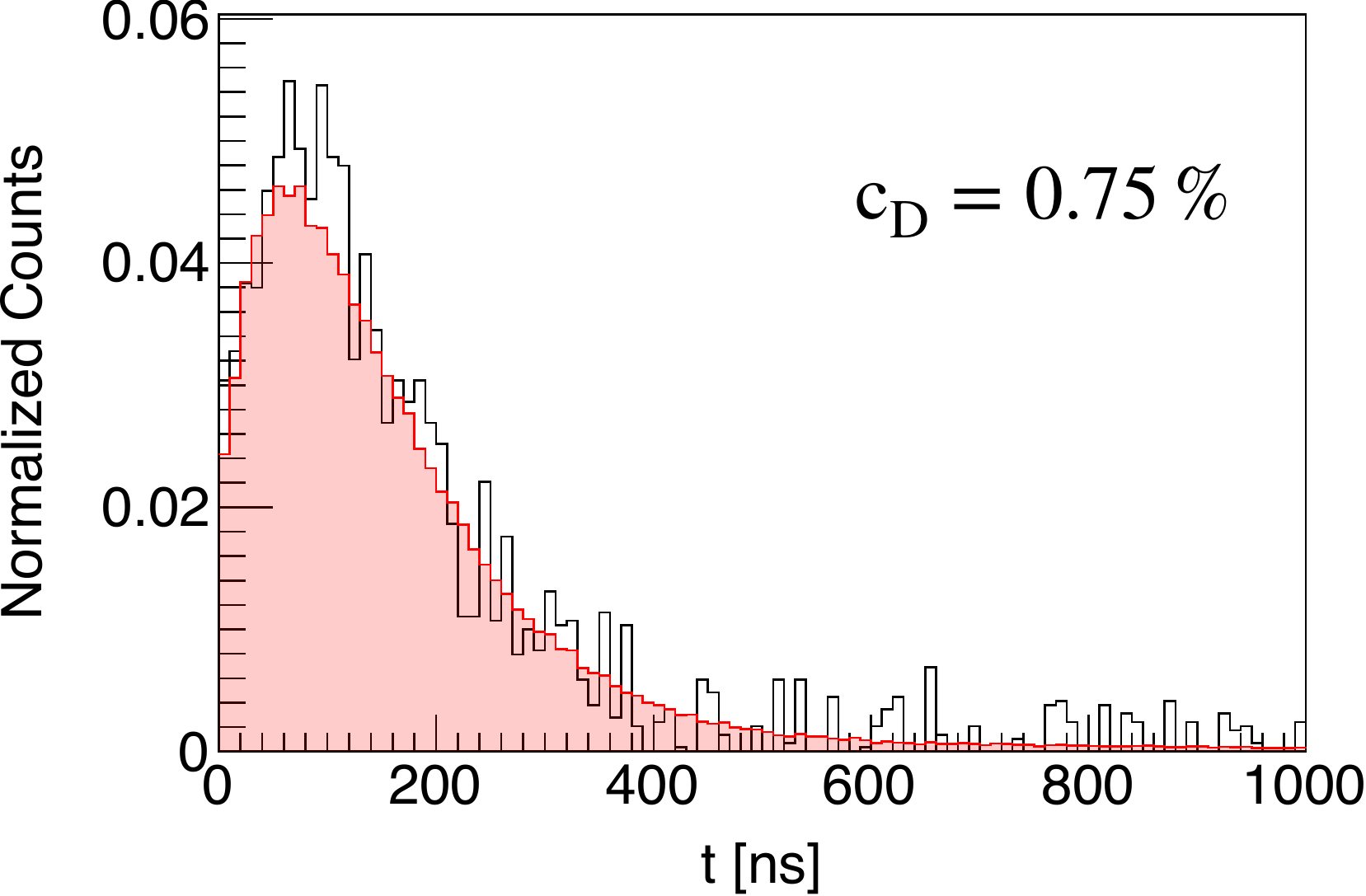} & 
\includegraphics[width=0.4\textwidth]{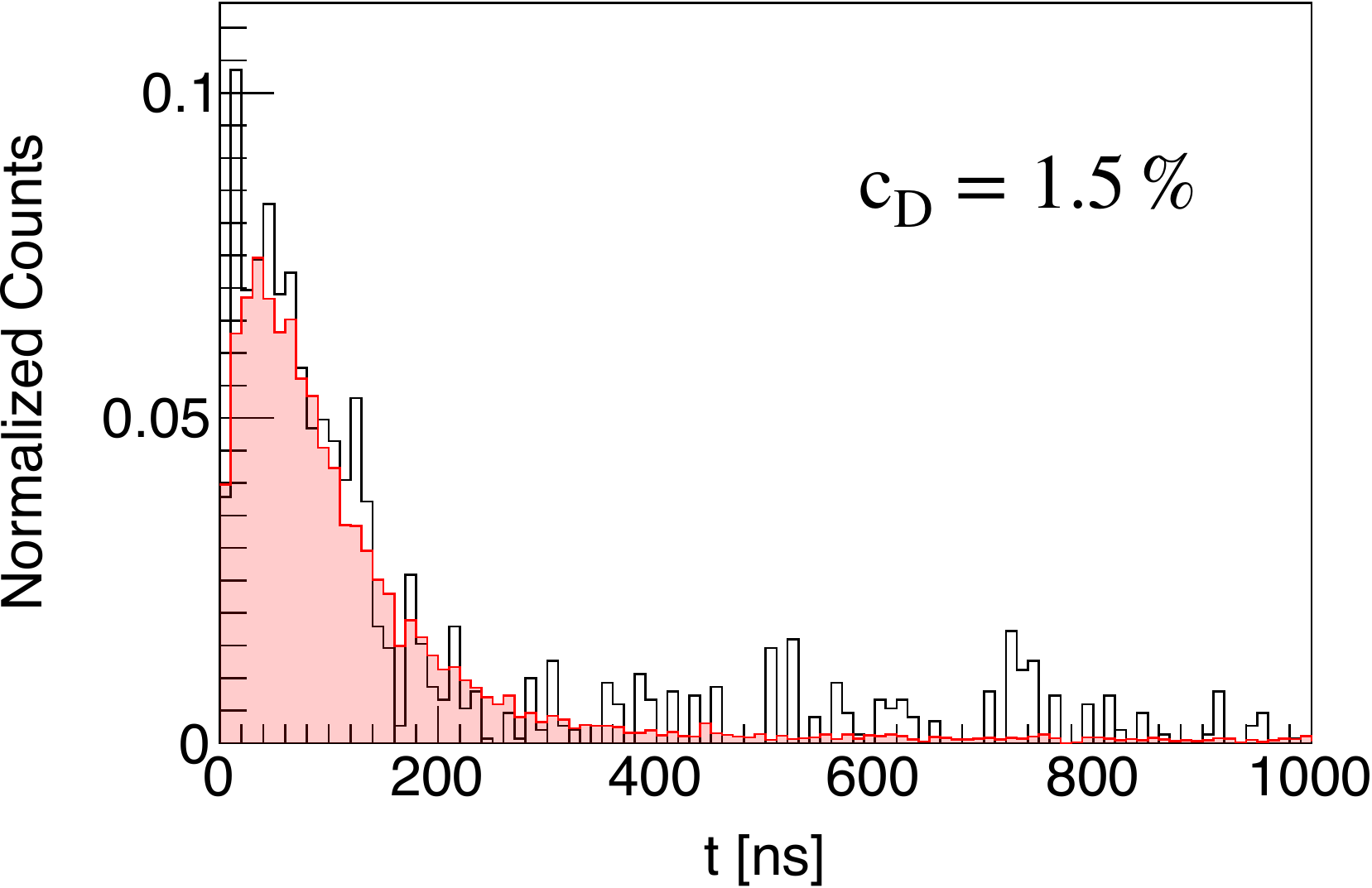} \\
(c) & (d) \\
\vspace{0.1cm} 
\end{tabular}
\caption{Evolution of the normalized muonic gold $2p-1s$ x rays as a function of time after a muon entered the gas cell for different deuterium concentrations: (a) $c_D=0.1\%$, (b) $c_D=0.25\%$, (c) $c_D=0.75\%$, and (d) $c_D=1.5\%$. The results from the simulation (in red) nicely agree with the measured data (in black).} \label{fig_time_distributions}
\end{figure*}

In order to measure the detection efficiency of the germanium array at the energies of the muonic gold $2p-1s$ x rays, a 300~\si{\micro\metre} thick natural lead target with a diameter of 30~mm was mounted in vacuum and without the entrance window and support grids at the target position inside the gas cell. This arrangement guaranteed that any muon registered by the entrance detector hits the lead target and stops within. By using the measured yield of the $2p-1s$ transition $\mathrm{BR}(2p-1s) = 0.899$ (from gold \cite{Har82} instead of lead as the corresponding values found in literature for lead \cite{But76, And69} of $\mathrm{BR}(2p-1s) \sim 0.6$ are clearly not correct and as only a negligible change of the $2p-1s$ yields are expected for these two close, high-$Z$ elements), the efficiency can be calculated by the ratio of detected number of x rays to the number of emitted x rays given by the number of muons multiplied by the yield: 
\begin{equation}
\epsilon_{det} = \frac{N_\gamma}{N_\mu \times \mathrm{BR}(2p-1s)}
\end{equation}
The number of detected x rays $N_\gamma$ was obtained by fitting the corresponding $2p-1s$ lines in the measured spectrum. Additionally and by taking full advantage of the large germanium detector array, a $\gamma-\gamma$ coincidence analysis was performed by triggering one of the germanium detectors on the $3d-2p$ transition and counting the number of observed $2p-1s$ x rays in the rest of detector array. The two approaches resulted in perfectly consistent values for the detector efficiency of $\epsilon_{det} = (0.46 \pm 0.01)\%$ at $\sim$5.6~MeV energy.

The window transmission $\zeta$ was measured just following the detector efficiency measurements by still employing the natural lead target, but closing the cell with the carbon fiber window and support grids and comparing the amount of emitted lead x rays per incoming muon with and without the entrance window. The extracted value amounts to $\zeta = 54.5\%$ and is consistent across several of the muonic lead lines.

With these parameters at hand the transfer efficiency can be calculated and plotted for the 2019 measurements as a function of deuterium concentration and beam momentum in Fig.~\ref{fig_scan_cd_p}, which reaches a maximum of approximately 5\% at a deuterium concentration of 0.25\% and a momentum of 27.25~MeV/c. The plots show the results for two different time windows of 500 and 1000~ns after the muon entrance highlighting the effect of the deuterium concentration and momentum on the speed of the transfer process to the gold target. Shown in addition to the measurements are also the results of the simulation from Fig.~\ref{fig_p_cD_Scan_2019}. Excellent agreement is found between the measurements and simulation suggesting good understanding of the processes occurring in the gas cell and corroborating the assumption that a 50~nm gold layer is thick enough such that all muons transfer to the target material. Good agreement between simulation and measurements is also found by plotting the time evolution of the muonic gold $2p-1s$ x rays as a function of time after a muon entered the gas cell -- see Fig.~\ref{fig_time_distributions}.

\subsection{Measurements with a few microgram target}
\begin{figure}%
\centering
\begin{tabular}{c}
\includegraphics[width=0.45\textwidth]{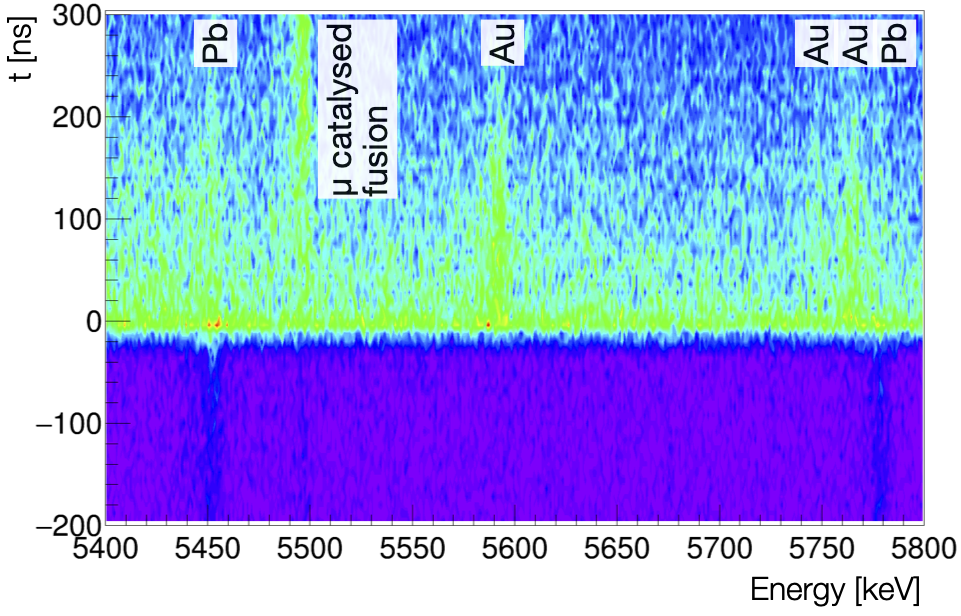} \\
(a) \\
\includegraphics[width=0.45\textwidth]{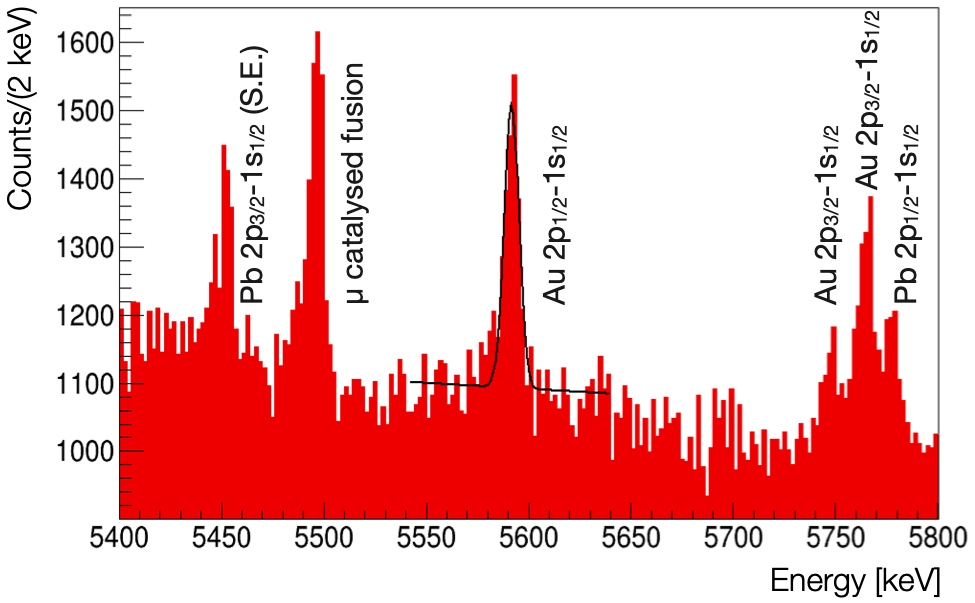} \\
(b)
\end{tabular}
\caption{Two dimensional histogram of the measured x-ray energy versus the time after a muon entered the gas cell (a) and its projection on energy (b). Visible are the $2p_{1/2}-1s_{1/2}$ and the single escape (S.E.) $2p_{3/2}-1s_{1/2}$ lines from the continuous muonic $^{208}$Pb calibration source, a gamma line from $p\mu d$ fusion and the muonic  $2p_{1/2}-1s_{1/2}$ and  $2p_{3/2}-1s_{1/2}$ x rays from the 5~\si{\micro\gram} gold target. The $2p_{3/2}-1s_{1/2}$ in muonic gold shows additionally a hyperfine structure. The black line in (b) is a simple Gaussian fit with linear background to the $2p_{1/2}-1s_{1/2}$ line of muonic gold. For details, see text.}
\label{fig_gold}
\end{figure}

Already in 2017 in a first proof-of-principle measurement, we have aimed at a measurement with only 5~\si{\micro\gram} of gold as motivated by the maximum allowed quantity for a potential $^{226}$Ra measurement. We did this in a series of measurements where the target mass was reduced from about 470~\si{\micro\gram} down to 5~\si{\micro\gram}. As we only discovered the issues with the hydrogen/deuterium mixture (as described in Sec.~\ref{sec_gas_system}) later, we cannot give precise values for these series of measurements as they are plagued with uncertainty due to changes in the deuterium concentration for the different measurements, but will only give some rough estimates in the next paragraph.  

In the measurement sequence we started with a 50~nm gold coating on a diameter of 25~mm (the target mount was somewhat different in these first measurements compared to the one described in Sec.~\ref{sec_target}) and gradually reduced it to a 3~nm coating on a $10\times 10$~mm$^2$ area. Each of the reductions -- 50~nm to 3~nm and 491~mm$^2$ to 100~mm$^2$ -- were accompanied by a loss in gold x rays of about a factor 3. For the final gold measurement with the 5~\si{\micro\gram} target material (3~nm coating on a $10\times 10$~mm$^2$), the estimated total transfer efficiency was slightly above 1\%. The result of a 18.5~h long measurement with this target is shown in Fig.~\ref{fig_gold}. The figure shows the measured energy spectrum versus the time after a muon entered the gas cell (Fig.~\ref{fig_gold}a) and its projection on energy (Fig.~\ref{fig_gold}b). Clearly visible are the peaks from the muonic $^{208}$Pb x rays used for the continuous calibration (see Sec.~\ref{sec_calibration}) that appear constant in time and the gold x rays from the target that only start to appear with the muon entering the gas cell. Additionally, the 5.5~MeV gamma ray emitted in the $p\mu d$ fusion process can be seen \cite{Fri91}.

Figure~\ref{fig_gold} also shows a simple Gaussian fit with linear background to the $2p_{1/2}-1s_{1/2}$ line of muonic gold. The resulting mean of the Gaussian is fitted to 5591.65(26)~keV perfectly in agreement and with only slightly worse precision than the value found in literature of 5591.71(15)~keV \cite{Fri04} demonstrating the excellent performance of the apparatus and method for such microgram target quantities. Based on this measured energy of the $2p_{1/2}-1s_{1/2}$, one could proceed to follow the standard procedure in muonic atom spectroscopy to extract a nuclear charge radius through the matching of the measured energy with calculations that have the nuclear charge radius as a free parameter \cite{Fri04}.

With the dominant uncertainty on any extracted, high-$Z$ nuclear charge radius being of theoretical nature, a measurement as shown in Fig.~\ref{fig_gold} that results in transition energies of the muonic atom with  $\sim$200~eV accuracy is sufficient to extract the charge radius of such high-$Z$ elements with a relative uncertainty of a few times $10^{-4}$ \cite{Fri04}.

\section{Conclusion}
In conclusion, we have developed a method and built an apparatus to perform muonic atom spectroscopy with targets that are only available in microgram quantities such as highly radioactive or scarce isotopically pure elements. The method relies on repeated transfer reactions taking place inside a 100~bar hydrogen gas cell with an admixture of 0.25\% deuterium at room temperature. The apparatus consists of a muon entrance detector, beam and decay electron veto detectors, a high-pressure gas cell containing the target and a high-purity germanium detector array to measure the energies of the x rays emitted in the muonic cascade.

In order to optimize and understand the apparatus, a detailed simulation was developed that models all of the processes taking place inside the gas cell. A comparison of the results of the simulation with the measured data shows excellent agreement and underlines the good understanding of the transfer processes taking place in such environments.

As a proof of principle, the muonic $2p-1s$ x rays from a 5~\si{\micro\gram} gold target were measured showing good signal to background levels and demonstrating the ability to perform high sensitivity measurements with microgram target materials. Employing this method, we succeeded to measure $^{248}$Cm, the heaviest element ever probed with muonic atom spectroscopy \cite{sps22}. A measurement of $^{226}$Ra is also within reach. Additional measurements of elements that were so far inaccessible are being planned.

\backmatter

%\bmhead{Supplementary information}
%If your article has accompanying supplementary file/s please state so here. 
%Please refer to Journal-level guidance for any specific requirements.

\bmhead{Acknowledgments}
%Acknowledgments are not compulsory. Where included they should be brief. Grant or contribution numbers may be acknowledged.
%Please refer to Journal-level guidance for any specific requirements.
We gratefully acknowledge fruitful discussions with P. Kammel that laid the basis to the development of the method presented in this paper. The experiments were performed at the $\pi$E1 beam line of PSI. We would like to thank the accelerator and support groups for the excellent conditions. Additional technical support by F. Barchetti, F. Burri, M. Hildebrandt, M. Meier, L. Noorda, A. Stoykov, and A. Weber from PSI is gratefully acknowledged. 

Funding was provided by FWO-Vlaanderen (Belgium), the Paul Scherrer Institute through the Career Return Program, the Cluster of Excellence ``Precision Physics, Fundamental Interactions, and Structure of Matter'' (PRISMA EXC 1098 and PRISMA+ EXC 2118/1) funded by the German Research Foundation (DFG) within the German Excellence Strategy (Project ID 39083149), the German Research Foundation (DFG) under Project WA 4157/1 and 407008443, the German BMBF under contracts 05P18PKCIA and `Verbundprojekt' 05P21PKCI1, and the Swiss National Science Foundation through the Marie Heim-V\"ogtlin Grant no. 164515 and project Grant no. 200021\_165569. P.I. is a member of the Allianz Program of the Helmholtz Association, contract no EMMI HA-216 ``Extremes of Density and Temperature: Cosmic Matter in the Laboratory''.

%\section*{Declarations}
%Some journals require declarations to be submitted in a standardised format. Please check the Instructions for Authors of the journal to %which you are submitting to see if you need to complete this section. If yes, your manuscript must contain the following sections under the %heading `Declarations':

%\begin{itemize}
%\item Conflict of interest/Competing interests (check journal-specific guidelines for which heading to use)
%\item Ethics approval 
%\item Consent for publication
%\item Availability of data and materials
%\item Authors' contributions
%\end{itemize}

%\noindent
%If any of the sections are not relevant to your manuscript, please include the heading and write `Not applicable' for that section. 


%\bibliography{bibliography}% common bib file


\end{document}